\documentclass[usenatbib]{mnras}
\usepackage{aas_macros}
\usepackage{amsfonts}
\usepackage{amsmath}
\usepackage{amssymb}
\usepackage{graphicx}
\usepackage[dvipsnames]{xcolor}
\usepackage{microtype}

\hypersetup{colorlinks=true,allcolors=teal}

\newcommand{\Mh}{\ensuremath{h^{-1}M_{\odot}}}
\newcommand{\Mhsq}{\ensuremath{h^{-2}M_{\odot}}}
\newcommand{\Mpch}{\ensuremath{h^{-1}{\rm Mpc}}}
\newcommand{\kpch}{\ensuremath{h^{-1}{\rm kpc}}}

\newcommand{\Nranmax}{\ensuremath{2\times10^8}}

\newcommand{\avg}[1]{\ensuremath{\left\langle \,#1\, \right\rangle}}

\newcommand{\der}{\ensuremath{{\rm d}}}

\newcommand{\mb}[1]{\ensuremath{\mathbf{#1}}}

\newcommand{\eqn}[1]{equation~\eqref{#1}}
\newcommand{\eqns}[1]{equations~\eqref{#1}}
\newcommand{\ph}[1]{\phantom{#1}}

\newcommand{\be}{\begin{equation}}
\newcommand{\ee}{\end{equation}}
\newcommand{\Cal}[1]{\ensuremath{\mathcal{#1}}}

\title[Voronoi volume function]
      {Voronoi volume function: A new probe of cosmology and galaxy evolution} 
\date{draft}

\author[Paranjape \& Alam]{
Aseem Paranjape$^{1}$\thanks{E-mail: aseem@iucaa.in}
\& Shadab Alam$^{2}$\thanks{E-mail: salam@roe.ac.uk} 
\\  
 $^1$ Inter-University Centre for Astronomy \& Astrophysics,
      Ganeshkhind, Post Bag 4, Pune 411007, India\\
 $^2$ Institute for Astronomy, University of Edinburgh, 
      Royal Observatory, Blackford Hill, Edinburgh, EH9 3HJ, UK
}

\begin{document}

\label{firstpage}
\pagerange{\pageref{firstpage}--\pageref{lastpage}}

\maketitle 

\begin{abstract}
\noindent
We study the Voronoi volume function (VVF) -- the distribution of cell volumes 
(or inverse local number density)
in the Voronoi tessellation of any set of cosmological tracers (galaxies/haloes).
We show that the shape of the VVF of biased tracers responds sensitively to physical properties such as halo mass, large-scale environment, substructure and redshift-space effects, making this a hitherto unexplored probe of both primordial cosmology and galaxy evolution. 
Using convenient summary statistics -- the width, median and a low percentile of the VVF as functions of average tracer number density -- we explore these effects for tracer populations in a suite of $N$-body simulations of a range of dark matter models. 
Our summary statistics sensitively probe primordial features such as small-scale oscillations in the initial matter power spectrum (as arise in models involving collisional effects in the dark sector), while being largely insensitive to a truncation of initial power (as in warm dark matter models). 
For vanilla cold dark matter (CDM) cosmologies, the summary statistics display strong evolution and redshift-space effects, and are also sensitive to cosmological parameter values for realistic tracer samples. 
Comparing the VVF of galaxies in the GAMA survey with that of abundance matched CDM (sub)haloes tentatively reveals environmental effects in GAMA beyond halo mass (modulo unmodelled satellite properties). 
Our exploratory analysis thus paves the way for using the VVF as a new probe of galaxy evolution physics as well as the nature of dark matter and dark energy. 
\end{abstract}


\section{Introduction}
\label{sec:intro}
\noindent
The spatial distribution of galaxies and inter-galactic gas in the cosmic web is the primary observable in furthering our understanding of the formation and evolution of large-scale structure in the Universe. This distribution of cosmic tracers of the dark matter field is affected not only by the non-linear physical processes that influenced the formation and evolution of the tracers, but also by primordial, cosmological variables, including the nature of dark matter and dark energy. As such, extracting and parsing the information content of this distribution has been and continues to be an exercise of great interest for both cosmology and galaxy evolution (\citealp[see, e.g.][]{amendola+18,troster+19}; \citealp[for a review, see][]{bcgs02}).

Among the many techniques employed in this endeavour is the \emph{Voronoi tessellation}, one of the oldest known methods for characterising the spatial distribution of a collection of points \citep{dirichlet1850,voronoi1908} and the focus of this work. The Voronoi tessellation, which gives a unique partitioning of space for a chosen set of discrete spatial points, has a long history of applications in a variety of scientific fields \citep[e.g.,][]{thiessen1911,gilbert62,kiang66,miles70,moller89} apart from cosmology and large-scale structure (\citealp[e.g.,][]{ivdw87,yi89}; \citealp[see also][and references therein]{vdw94}). Given a discrete set of tracer locations (seeds, or nuclei) in space, the Voronoi tessellation partitions space by assigning to each tracer a cell containing those points which are closer to this tracer than to any other tracer. 

The Voronoi tessellation has played a key role in algorithms associated with cosmic web classification \citep{SchaapVandeWeygaert2000,aragoncalvo07}, cosmic velocity field reconstruction (\citealp{bvdw96}; see also \citealp{haa15}), void identification \citep{pvdwj07,Neyrinck2008}, cosmological hydrodynamical simulations \citep{springel10,vogelsberger+12}, etc. The spatial distribution of the \emph{vertices} of the Voronoi tessellation of randomly (i.e., Poisson) distributed spatial points also formed the basis of some early models of galaxy clustering \citep{vdwi89,coles90,vdw91,ms02}. A comprehensive study of the mathematical properties of Voronoi tessellations of some specific examples of clustered point sets, including the distribution of cell volumes, faces, edges and vertices of these tessellations, can be found in \citet{vdw94}.

\begin{table*}
\centering
\begin{tabular}{cccccccc}
\hline
\hline
cosmology${}^{\,a}$ & dark matter & configuration${}^{\,c}$ & $m_{\rm part}$ & $\epsilon_{\rm f}$ (com.) & $z_{\rm init}$ & $N_{\rm real}$ & merger \\
& model${}^{\,b}$ &  & $(10^9\Mh)$ & $(\kpch)$ & & & trees \\
\hline
\hline
WMAP7 & CDM & ${\rm L}600\_{\rm N}1024$ & $15.41$ & $19.5$ & $99$ & $3$ & no \\
&  & ${\rm L}300\_{\rm N}1024$ & $1.93$ & $9.8$  & $49$ & $3$ & no \\
&  & ${\rm L}150\_{\rm N}1024$ & $0.24$ & $4.9$  & $99$ & $2$ & yes \\
& WDM  & ${\rm L}150\_{\rm N}1024$ & $0.24$ & $4.9$ & $99$ & $1$ & yes \\
\hline
P13 & CDM  & ${\rm L}150\_{\rm N}512\ph{0}$ & $2.20$ & $9.8$ & $49$  & $1$ & no \\
\hline
P18 & CDM  & ${\rm L}200\_{\rm N}1024$ & $0.63$ & $6.5$ & $99$ & $1$ & yes \\
& BDM  & ${\rm L}200\_{\rm N}1024$ & $0.63$ & $6.5$ & $99$ & $1$ & yes \\
\hline
\hline
\end{tabular}
\caption{{\bf Summary of simulations used in this work:} Columns respectively indicate cosmology, dark matter model, simulation configuration (see notes), particle mass $m_{\rm part}$ in $10^9\Mh$, comoving force softening scale $\epsilon_{\rm f}$ in \kpch, initial redshift $z_{\rm init}$, number of realisations used $N_{\rm real}$ and whether or not merger trees were available. \underline{\emph{Notes:}} (a) All models assumed a flat background cosmology with parameters $\{\Omega_{\rm m},\Omega_{\Lambda},\Omega_{\rm b},h,n_{\rm s},\sigma_8\}$ given by $\{0.276,0.724,0.045,0.7,0.961,0.811\}$ {\bf (WMAP7)}, $\{0.315,0.685,0.049,0.673,0.96,0.829\}$ {\bf (P13)} and $\{0.306,0.694,0.0484,0.678,0.9677,0.815\}$ {\bf (P18)}. (b) CDM simulations used standard cold dark matter transfer functions to generate initial conditions, while WDM and BDM simulations used non-standard transfer functions with small-scale features as described in section~\ref{sec:cosmo}. (c) The configuration denotes the combination of box size $L_{\rm box}$ and particle number $N_{\rm part}$ as, e.g., ${\rm L}150\_{\rm N}1024$ for $L_{\rm box}=150\Mpch$, $N_{\rm part}=1024^3$.}
\label{tab:sims}
\end{table*}

In this work, we explore the cosmological information content of an easily measurable but hitherto neglected observable, namely, the \emph{distribution of Voronoi volumes of a set of clustered tracers} of the dark matter field (dark haloes or galaxies), as a function of tracer properties. We will do so from the point of view of both cosmology as well as galaxy evolution. Although this observable has been previously discussed in the literature \citep[e.g.,][]{vdw94}, to the best of our knowledge its cosmological information content, particularly in the context of realistic biased tracers of dark matter, has not been systematically explored \citep[see, however,][for studies of the VVF of dark matter particles in $N$-body simulations]{Neyrinck2008,neyrinck13,yang+15}.

Theoretically, as we will show below, the shape of this `Voronoi volume function' (VVF) is closely connected to the void probability function \citep{fgjw76,white79a} of the given tracer set, which measures the probability of a randomly placed region (typically, a sphere of fixed radius) to be empty of all tracers. This is not surprising, since each Voronoi cell is a region that is empty of all except one tracer. The void probability function contains contributions from the entire infinite hierarchy of spatial $N$-point correlation functions of the tracer field \citep{white79a}, and the VVF consequently inherits this treasure trove of non-linear cosmological information. 
Since Voronoi cells, by construction, are subject to more constraints with regards their shape and the positioning of neighbouring cells than are randomly distributed empty regions, we expect the non-linear information in the VVF to be packaged very differently than in the void probability function.
We will use $N$-body simulations to explore the nature of the VVF as a function of a variety of tracer properties such as mass, clustering strength, substructure content and redshift space effects.

From the computational point of view, several efficient algorithms have been developed for generating the Voronoi tessellation and extracting its properties such as the number and distribution of vertices, faces, edges, etc. (\citealp[e.g.,][]{meijering53,boots74,bdf78,ivdw87,obs92,vdw94}; \citealp[see also][and references therein]{fn07}). Below we will discuss a simple Monte Carlo technique for estimating Voronoi cell volumes (the sole focus of our work) which is easily and robustly extendable to galaxy catalogs affected by masking and incompleteness. 

The paper is organised as follows. In section~\ref{sec:sims}, we describe our $N$-body simulations. In section~\ref{sec:vvf:theory}, we discuss theoretical aspects of the distribution of Voronoi volumes, emphasising its sensitivity to the infinite hierarchy of tracer correlation functions. Section~\ref{sec:vvf:sims} presents our numerical results in cold dark matter (CDM) cosmologies, for haloes and subhaloes selected by mass and clustering strength, in real and redshift space. In section~\ref{sec:GAMA}, we attempt to construct a tracer sample whose VVF matches what is observed for luminosity-thresholded galaxy samples in the GAMA survey. We explore the dependence of Voronoi volumes on dark matter properties in section~\ref{sec:cosmo} using simulations of two non-standard dark matter models, and we conclude in section~\ref{sec:conclude}. The Appendices discuss a few technical aspects of our  analysis: Appendix~\ref{app:negbin} gives some details on the void probability function, Appendix~\ref{app:downmaskassembly} calibrates the dependence of the Voronoi volume statistics on selection effects due to downsampling, survey masks and halo assembly bias, and Appendix~\ref{app:hbyhbias} gives details of our choice of estimator for the large-scale clustering of individual tracers.

For most of the analysis, we use a spatially flat Lambda cold dark matter ($\Lambda$CDM) cosmology and explore various values of total matter density parameter $\Omega_{\rm m}$, baryonic matter density $\Omega_{\rm b}$, Hubble constant $H_0=100h\,{\rm kms}^{-1}{\rm Mpc}^{-1}$, primordial scalar spectral index $n_{\rm s}$ and r.m.s. linear fluctuations in spheres of radius $8\Mpch$, $\sigma_8$, with transfer functions generated by the codes \textsc{camb} \citep{camb}\footnote{\href{http://camb.info}{http://camb.info}} and \textsc{class} \citep{class-I,class-II}.\footnote{\href{http://class-code.net}{http://class-code.net}} In section~\ref{sec:cosmo} we explore two alternate collisionless dark matter models affected by small-scale features involving, respectively, a truncation and oscillations in the initial matter power spectrum.

\vskip 0.5in
\section{Simulations}
\label{sec:sims}
\noindent
We use a suite of simulations spanning multiple cosmological parameter sets, box sizes and mass resolutions. Table~\ref{tab:sims} and its caption summarise these details. We use three sets of cosmological parameters denoted WMAP7, P13 and P18, respectively compatible with the 7-year results of the \emph{Wilkinson Microwave Anisotropy Probe} experiment \citep{Komatsu2010} and the \emph{Planck} experiment's results from 2013 \citep{Planck13-XVI-cosmoparam} and 2018 \citep{Planck18-VI-cosmoparam}. The parameter values for each of these are given in the caption of Table~\ref{tab:sims}.

In the following, we denote each simulation configuration by its cosmology and the combination of (periodic) box size $L_{\rm box}$ and particle number $N_{\rm part}$ as, e.g., ${\rm L}150\_{\rm N}1024$ for $L_{\rm box}=150\Mpch$, $N_{\rm part}=1024^3$. For several configurations, we used multiple realisations performed after changing the random number seed used for generating the initial conditions. For two of the configurations, namely WMAP7 ${\rm L}150\_{\rm N}1024$ and P18 ${\rm L}200\_{\rm N}1024$, we have one paired simulation each (i.e., performed with the same random seed for the initial conditions, respectively) using non-standard dark matter models which we describe in more detail in section~\ref{sec:cosmo} below. 

\begin{figure}
\centering
\includegraphics[width=0.45\textwidth]{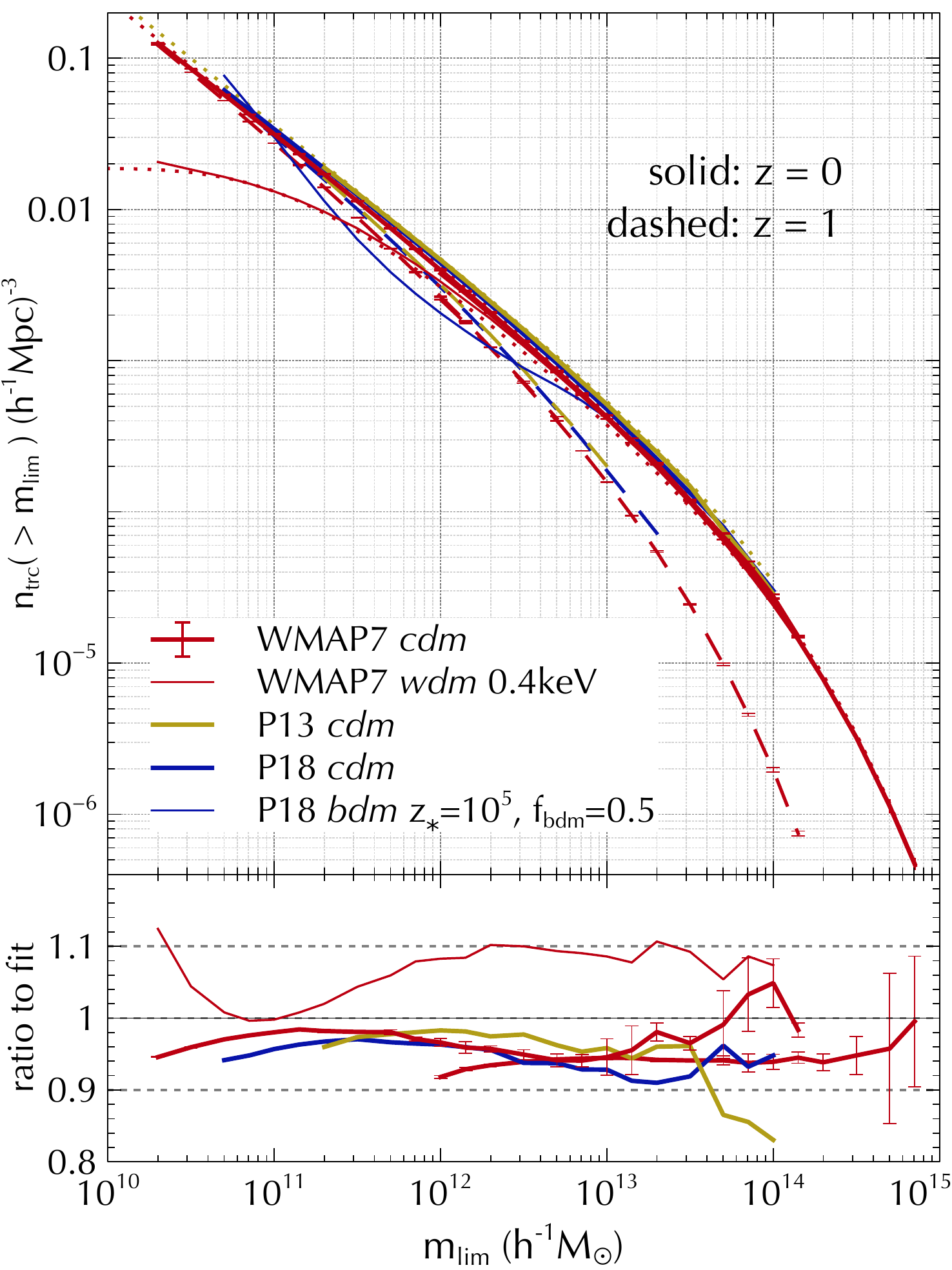}
\caption{{\bf Halo mass functions:} \emph{(Top panel:)} Cumulative comoving number density of haloes as a function of mass threshold (using the $m_{\rm 200b}$ definition, see text). Red, yellow and blue curves correspond to the WMAP7, P13 and P18 cosmological parameter sets, respectively. Thick curves correspond to CDM configurations and thin curves to the WDM and BDM models described in section~\ref{sec:cosmo}. Solid (dashed) curves show results at redshift $z=0$ ($z=1$). Dotted curves show fitting functions for the CDM \citep{Tinker08} and WDM cases \citep{ssr13}. Such fits are unavailable for the BDM case at present. \emph{(Bottom panel:)} Ratio of the CDM and WDM measurements at $z=0$ with the appropriate fits from the top panel. The two sets of WMAP7 CDM results are for the configurations ${\rm L}150\_{\rm N}1024$ (extending to lower masses) and ${\rm L}600\_{\rm N}1024$ (higher masses). Both of these were averaged over all available realisations (see Table~\ref{tab:sims}), with error bars indicating the standard deviation across realisations.}
\label{fig:hmf}
\end{figure}

The simulations were performed using the tree-PM code \textsc{gadget-2} \citep{springel:2005}\footnote{\href{http://www.mpa-garching.mpg.de/gadget/}{http://www.mpa-garching.mpg.de/gadget/}} with a PM grid of a factor $2$ finer than the initial particle count along each axis, and a comoving force softening length of $1/30$ of the mean interparticle spacing. The particle masses and softening lengths for each of our configurations are summarised in Table~\ref{tab:sims}. Initial conditions were generated using $2^{\rm nd}$ order Lagrangian perturbation theory \citep{scoccimarro98} with the code \textsc{music}  \citep{hahn11-music}.\footnote{\href{https://www-n.oca.eu/ohahn/MUSIC/}{https://www-n.oca.eu/ohahn/MUSIC/}} Haloes were identified in each box using the code \textsc{rockstar} \citep{behroozi13-rockstar}\footnote{\href{http://code.google.com/p/rockstar/}{http://code.google.com/p/rockstar/}} which implements a Friends-of-Friends algorithm in 6-dimensional phase space. For the configurations WMAP7 ${\rm L}150\_{\rm N}1024$ and P18 ${\rm L}200\_{\rm N}1024$, we stored $201$ snapshots equally spaced in the scale factor $a=1/(1+z)$ ($\Delta a=0.004615$) between $z=12$ and $z=0$, which we used to produce merger trees using the code \textsc{consistent-trees} \citep{behroozi13-consistenttrees}.\footnote{\href{https://bitbucket.org/pbehroozi/consistent-trees}{https://bitbucket.org/pbehroozi/consistent-trees}} We only retain relaxed haloes whose virial ratio $\eta=2T/|U|$ satisfies $0.5\leq\eta\leq1.5$ as prescribed by \cite{bett+07}. Unless stated otherwise, we will quote halo masses using the definition $m_{\rm 200b}$ which denotes the gravitationally self-bound mass contained in the radius $R_{\rm 200b}$ at which the enclosed dark matter density becomes $200$ times the mean density of the universe. We will mostly focus on results at $z=0$ in this work. All simulations were performed on the Perseus cluster at IUCAA.\footnote{\href{http://hpc.iucaa.in}{http://hpc.iucaa.in}}

Figure~\ref{fig:hmf} shows the cumulative number density of haloes as a function of mass for various cosmologies and redshifts. All cosmologies, including the non-standard dark matter models, behave similarly at high masses, showing the well-understood exponential cutoff in number counts \citep{ps74,bcek91,st99}. At lower masses, the non-standard dark matter models depart from the power law behaviour characteristic of CDM cosmologies. We will return to these features in section~\ref{sec:cosmo}; for now, we simply note from Figure~\ref{fig:hmf} that our simulations span a wide variety of behaviours of the halo mass function.

\section{Voronoi volume function: Theory}
\label{sec:vvf:theory}
\noindent
Given $N_{\rm trc}$ tracers spread over a region of volume $V_{\rm tot}$, we are interested in the distribution of the quantity $y$ defined by
\be
y \equiv V / \avg{V} = n_{\rm trc}V\,,
\label{eq:y-def}
\ee
where $V(t)$ is the volume of the Voronoi cell of an individual tracer $t$ and $\avg{V}$ is the mean volume which is simply the inverse of the tracer number density $n_{\rm trc}$:
\be
\avg{V} = \frac1{N_{\rm trc}} \sum_{t=1}^{N_{\rm trc}}\,V(t) = \frac{V_{\rm tot}}{N_{\rm trc}} = n_{\rm trc}^{-1}\,.
\label{eq:<V>}
\ee
Below, we first recall the known theoretical properties of the distribution $p(y)$ for a set of \emph{unclustered} (or Poisson distributed) tracers in 3 dimensions, before discussing the more relevant case of clustered tracers. The Poisson case will serve as a useful reference throughout the analysis. We will refer to the distribution $p(y)$ as the Voronoi volume function (henceforth, VVF). By construction, we have 
\be
\avg{y} = \int\der y\,p(y)\,y = 1\,,
\label{eq:<y>}
\ee
for clustered as well as unclustered tracers.

\subsection{Poisson distributed tracers}
\label{subsec:theory:Poisson}
\noindent
For unclustered tracers in 3 dimensions, the second moment $\avg{y^2}$ of the VVF has an exact analytical expression \citep[equations 10, 14 and Table II of][see also below]{gilbert62}
\begin{align}
\avg{y^2}_{\rm Poisson} &= \frac{8\pi^2}{3}\int_0^\infty\der \zeta\,\zeta^2\int_{-1}^{1}\der\mu\,\frac1{V(\zeta,\mu)^2} \notag\\
&= 1.179\,,
\label{eq:Poisson<y^2>}
\end{align}
where $V(\zeta,\mu)$ is given by
\begin{align}
V(\zeta,\mu) &= \frac\pi3\bigg[2\zeta^3 + 3\mu \zeta(\zeta^2+1) - (3\mu^2\zeta^2+1) \notag\\
&\ph{\pi/3[]+}
+ 3(1-\mu \zeta)\left|\zeta^2+1-2\mu \zeta\right| \notag\\
&\ph{\pi/3[]+3()}
+ 2\left|\zeta^2+1-2\mu \zeta\right|^{3/2}\bigg]\,,
\label{eq:V(zeta,mu)}
\end{align}
and the integrals in the first line of \eqn{eq:Poisson<y^2>} must be performed numerically.

Although there are no corresponding exact analytical results for the shape of the full distribution $p(y)$, this is known to be accurately described by a 3-parameter generalised Gamma function model \citep[see, e.g.,][]{wkw86,kumar+92,tanemura03,fn07},
\be
p_{\rm Poisson}(y) = \frac{c\,b^{a/c}}{\Gamma(a/c)}\,y^{a-1}{\rm exp}(-by^c)\,,
\label{eq:Poisson-p(y)}
\ee
where $\Gamma(x)$ is the Gamma function. In the following, when quoting results for $y$-percentiles of Poisson distributed tracers, we will adopt the values $a=4.8065$, $b=4.06342$ and $c=1.16391$ \citep{tanemura03}.\footnote{The values $a=3.24174$, $b=3.24269$ and $c=1.26861$ as reported by \citet[][]{fn07} appear to be erroneous since they do not reproduce Figure 6 of that paper.}

\subsection{Clustered tracers}
\label{subsec:theory:clustered}
\noindent
In this section, we sketch a formal derivation of the second moment $\avg{y^2}$ of the VVF of arbitrarily clustered tracers. We closely follow the treatment in \citet{gilbert62} combined with the formalism for the hierarchy of correlation functions developed by \citet{white79a}. This analysis will demonstrate that, similarly to the void probability function, the VVF contains information on the full clustering hierarchy of the tracer population. Subsequent sections will explore the empirical consequences of this fact.

\citet{gilbert62} recasts the second moment $\avg{y^2}$ of Poisson distributed tracers with number density $n_{\rm trc}$ (we will switch to clustered tracers momentarily) as the calculation of the average volume of the Voronoi cell which contains a specified, non-tracer point (taken to be the origin). Let this cell be associated with tracer $t$ located at $\mb{x}_{t}$. The required average volume can be written as an integral over $s$ of the probability $P(s)$ that a point $\mb{s}$ at distance $s$ from the origin belongs to this same Voronoi cell $t$. \citeauthor{gilbert62} shows that this leads, in 3 dimensions, to the expression
\begin{align}
\avg{y^2} &= 4\pi n_{\rm trc}\int_0^\infty\der s\,s^2\,P(s) \notag\\
&= 4\pi n_{\rm trc}\int_0^\infty\der s\,s^2\,(2\pi)n_{\rm trc}s^3 \notag\\
&\ph{4\pi n_{\rm trc}\int_0^\infty}
\times\int_0^\infty\der \zeta\,\zeta^2\int_{-1}^1\der\mu\,\exp(W_0)
\label{eq:<y^2>derivn}
\end{align}
where $\zeta$ is defined such that $|\mb{x}_t| = \zeta s$ and $\mu$ is the cosine of the angle between $\mb{s}$ and $\mb{x}_t$, $\mu = \mb{s}\cdot\mb{x}_t/(\zeta s^2)$. Consider two spheres, respectively centered at the origin and at \mb{s}, and each containing the tracer location $\mb{x}_t$ on their surface. \emph{The factor $\exp(W_0)$ in \eqn{eq:<y^2>derivn} is the probability that the union of these two spheres is empty.} If the volume of this union $V_U$ is written as $V_U = s^3V(\zeta,\mu)$, then \citeauthor{gilbert62} shows that $V(\zeta,\mu)$ is given by \eqn{eq:V(zeta,mu)}.

By construction, the factor $\exp(W_0)$ is the void probability function (henceforth, VPF) for the volume $V_U$. For the Poisson case, $W_0(n_{\rm trc},V_U)=-n_{\rm trc}V_U$, while $W_0(n_{\rm trc},V_U)$ for clustered tracers is an infinite sum over all $N$-point tracer correlation functions averaged over $V_U$ \citep{white79a,sheth96}:
\be
W_0(n_{\rm trc},V) = \sum_{k=1}^\infty\frac{\left(-n_{\rm trc}V\right)^k}{k!}\,\bar\xi_k(V) \equiv \left(-n_{\rm trc}V\right)\,\chi(n_{\rm trc},V)\,,
\label{eq:W0-def}
\ee
where $\bar\xi_1\equiv 1$ and
\be
\bar\xi_k(V) = \prod_{i=1}^k\left(\frac1V\int_V\der^3x_i\right)\,\xi_k\left(\mb{x}_1,\ldots,\mb{x}_k\right)\,,\quad k\geq2\,,
\label{eq:xibar_k}
\ee
with $\xi_k\left(\mb{x}_1,\ldots,\mb{x}_k\right)$ being the connected $k$-point correlation function of the tracer locations \citep[][so that $\xi_2(\mb{x}_1,\mb{x}_2)=\xi(|\mb{x}_1-\mb{x}_2|)$ is the usual 2-point correlation function]{peebles-1980,bcgs02}, and we introduced the `reduced VPF' $\chi(n_{\rm trc},V)$ in keeping with the general VPF literature: $\chi=1$ for Poisson tracers. Appendix~\ref{app:negbin} gives some further details regarding the observed shape of the VPF for galaxy samples.

\begin{figure*}
\centering
\includegraphics[width=0.48\textwidth]{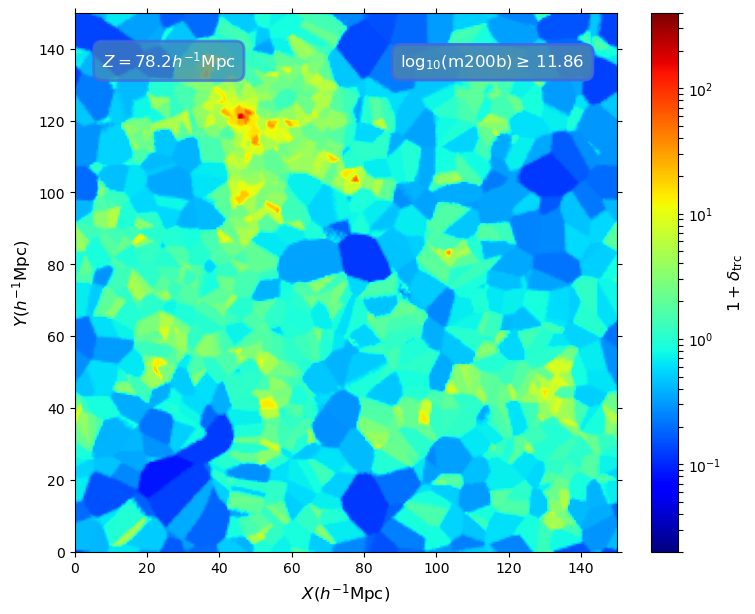}
\includegraphics[width=0.48\textwidth]{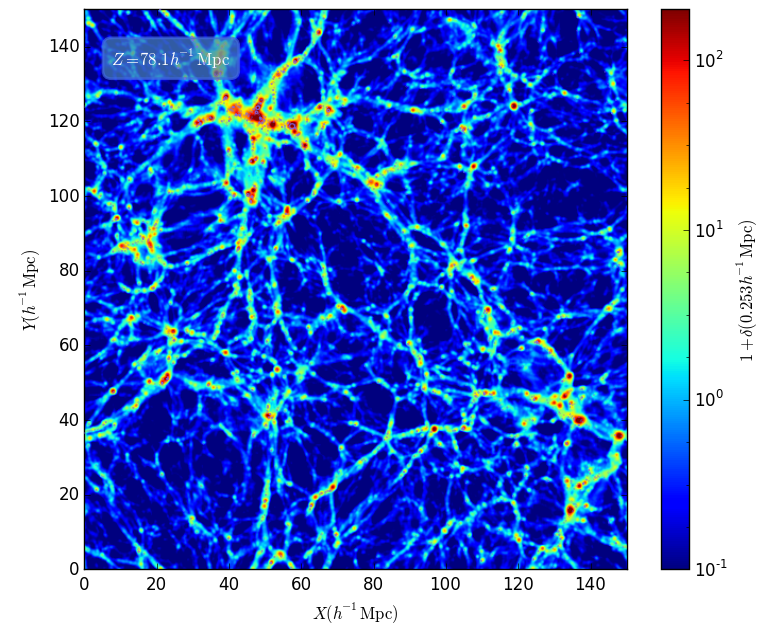}
\caption{{\bf Voronoi tessellation of simulated haloes:} \emph{(Left panel:)} Visualisation of the Voronoi cells of mass-selected haloes in one of the WMAP7 CDM ${\rm L}150\_{\rm N}1024$ boxes at $z=0$. The cell volumes $V$ were estimated using the Monte Carlo algorithm described in section~\ref{subsec:tessellation} and are coloured by the overdensity estimator $1+\delta_{\rm trc}=\left(n_{\rm trc}V\right)^{-1}$, where $n_{\rm trc}$ is the tracer number density of the box. The image shows a single-cell slice of this density field evaluated on a $256^3$ grid. \emph{(Right panel:)} Full dark matter density field in the same spatial slice, estimated using cloud-in-cell interpolation on a $512^3$ grid and smoothed with a Gaussian kernel of radius $\sim250\kpch$.}
\label{fig:vormap}
\end{figure*}

After some re-arrangement and transforming $s\to\bar N \equiv n_{\rm trc}V_U = n_{\rm trc}s^3V(\zeta,\mu)$ in \eqn{eq:<y^2>derivn}, the second moment $\avg{y^2}$ of the VVF can be written, in general, as
\begin{align}
\avg{y^2} &= \frac{8\pi^2}{3} \int_0^\infty\der \zeta\,\zeta^2\int_{-1}^1\der\mu\,\frac1{V(\zeta,\mu)^2} \notag\\
&\ph{\frac{8\pi^2}{3} \int}
\times\int_0^\infty\der\bar N\,\bar N\,\exp\left(-\bar N\chi(\bar N,\zeta,\mu)\right) \notag\\
&= 1.179 \avg{\int_0^\infty\der\bar N\,\bar N\,\exp(-\bar N\chi)}_{(\zeta,\mu)}\,,
\label{eq:<y^2>}
\end{align}
where, in the second line, we used \eqn{eq:Poisson<y^2>} and introduced the weighted average $\avg{f}_{(\zeta,\mu)}$ of some function $f(\zeta,\mu)$, 
 \be
\avg{f}_{(\zeta,\mu)} \equiv \frac{\int_0^\infty\der \zeta\,\zeta^2\int_{-1}^1\der\mu\,f(\zeta,\mu)/V(\zeta,\mu)^2}{\int_0^\infty\der \zeta\,\zeta^2\int_{-1}^1\der\mu\,1/V(\zeta,\mu)^2}\,.
\label{eq:(f)_(zeta,mu)}
\ee
As a check, note that the unclustered case $\chi=1$ leads to $\int_0^\infty\der\bar N\,\bar N\,\exp(-\bar N)=1$, so that \eqn{eq:<y^2>} recovers \eqn{eq:Poisson<y^2>}, as it should.

\section{Voronoi volume function: Simulations}
\label{sec:vvf:sims}
\noindent
We now explore the properties of the VVF of dark matter haloes selected according to various criteria, with the aim of identifying interesting features that may be particularly sensitive either to cosmology or non-linear physics. 
In this section, we focus on results for CDM cosmologies and explore the VVF in non-standard dark matter models in section~\ref{sec:cosmo}. 

\subsection{Generating the Voronoi tessellation}
\label{subsec:tessellation}
\noindent
We start by describing our method for generating the Voronoi tessellation of a given set of tracers.
As in the previous section, consider a set of $N_{\rm trc}$ tracers with 3-dimensional positions $\{\mb{x}_t\}$ with $1\leq t\leq N_{\rm trc}$, in a region of volume $V_{\rm tot}$. Our algorithm is the same as described by \citet{azpm19} for their density field reconstruction and proceeds as follows:

\begin{enumerate}
\item Generate $N_{\rm ran}\gg N_{\rm trc}$ uniform random positions $\{\mb{x}_r\}$ with $1\leq r\leq N_{\rm ran}$ in the full volume. (We discuss the value of $N_{\rm ran}$ below.)
\item For each random point $r$, find the nearest tracer $t$ and `assign' the random point to this tracer. This can be efficiently done using, e.g., KD trees.
\item Now for each tracer $t$, count the number of randoms $\nu_{\rm ran}(t)$ assigned to $t$. The volume of the Voronoi cell associated with the tracer $t$ is then estimated as 
\be
V(t) = V_{\rm tot}\nu_{\rm ran}(t)/N_{\rm ran}\,.
\label{eq:V(t)}
\ee
\end{enumerate}
Figure~\ref{fig:vormap} shows a visualisation of the resulting Voronoi tessellation using haloes selected by a mass threshold \emph{(left panel)} compared with the full dark matter density field of the box \emph{(right panel)}.

In a periodic box, all distances between the randoms and the tracers account for periodicity, leading to a unique tessellation. In a realistic survey, the survey boundary must also be accounted for, a complication we will ignore for the time being. We do note, however, that the Monte Carlo approach described above can naturally account for survey incompleteness, simply by downsampling the randoms in any given region according to its expected incompleteness.

Experimenting with the value of $N_{\rm ran}$, we found that individual Voronoi volumes $V$ converge at better than $\sim2\%$ (at the $1\sigma$ level) for $N_{\rm ran}\gtrsim3\times10^4\times N_{\rm trc}$. We are, however, more interested in the distribution of $V$, which is more robust to sampling errors. We therefore impose an upper limit $N_{\rm ran,max}$ to the total number of random points $N_{\rm ran}$ we generate in the full volume $V_{\rm tot}$ for any tracer population. We have found that all our results are well-converged for $N_{\rm ran,max}\gtrsim10^8$; as a conservative choice we display results for $N_{\rm ran,max}=\Nranmax$. 
For uniformity, we generate $N_{\rm ran,max}$ randoms for \emph{all} tracer populations. 

\subsection{Tracers selected by halo mass}
\label{subsec:massthreshold}

\begin{figure}
\centering
\includegraphics[width=0.45\textwidth]{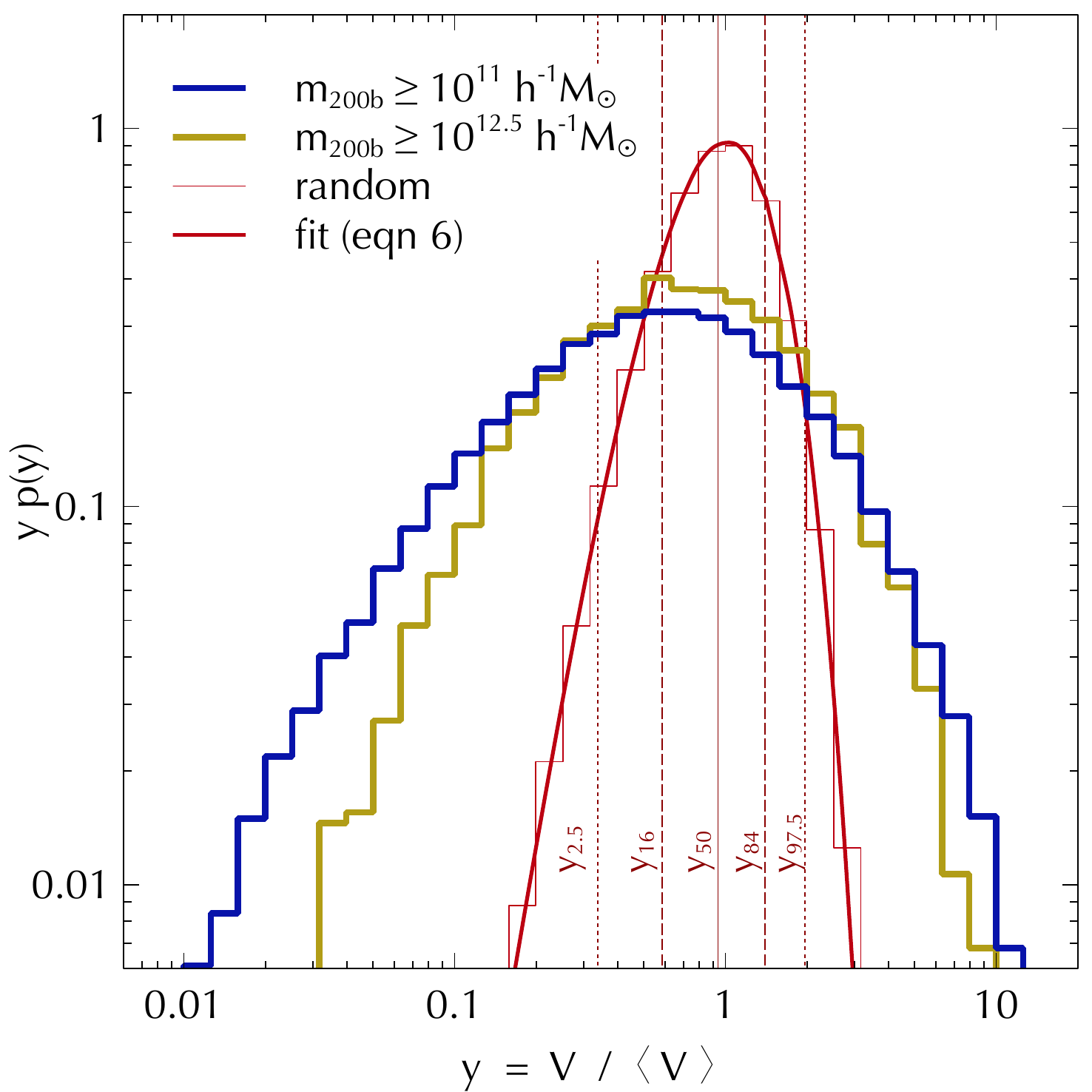}
\caption{{\bf Differential VVF} for real-space haloes selected by a mass threshold $m_{\rm 200b}>m_{\rm lim}$ at $z=0$ for the WMAP7 CDM cosmology, for two choices of $m_{\rm lim}$ (thick blue and yellow histograms), compared with the VVF of randomly distributed (Poisson) points in the same volume (thin red histogram). The latter is accurately described by the fitting function \eqref{eq:Poisson-p(y)} from the literature (smooth red curve). 
We indicate the percentiles $\{y_{2.5},y_{16},y_{50},y_{84},y_{97.5}\} = \{0.338,0.585,0.940,1.403,1.964\}$ of the Poisson VVF as vertical lines (computed using equation~\ref{eq:Poisson-p(y)}). The halo VVFs are clearly sensitive to $m_{\rm lim}$ and broader than the Poisson VVF. Being unimodal, the VVFs are fully described by their percentiles, which will be exclusively used in subsequent plots.}
\label{fig:vvfpdf}
\end{figure}

\noindent
As the simplest and most intuitive selection criterion, let us first study the shape of the VVF of haloes selected by a mass threshold $m_{\rm 200b} > m_{\rm lim}$. 
Figure~\ref{fig:vvfpdf} shows the differential VVF of two halo samples selected using $m_{\rm lim}=10^{11}\Mh$ and $10^{12.5}\Mh$, respectively, and shown as thick solid histograms (the higher threshold leads to a narrower distribution). We used the $z=0$ snapshot of one of the WMAP7 ${\rm L}150\_{\rm N}1024$ CDM boxes for selecting these samples. For comparison, the thin histogram shows the VVF of randomly distributed points in the same volume, which is accurately described by \eqn{eq:Poisson-p(y)} (smooth curve). We see that both the halo VVFs are unimodal and broader than the Poisson VVF. The latter is easily understood as a consequence of clustering: imagine `moving' an unclustered set of $N$ points into a configuration identical to the actual positions of $N$ haloes in a simulation volume $V_{\rm box}$. This would involve bringing together groups of these points so as to become clustered near filaments and nodes, while simultaneously emptying underdense voids. Clearly, this will increase the number of, both, small-volume as well as large-volume cells, while keeping the mean cell volume intact at $V_{\rm box}/N$, which is equivalent to broadening the VVF. 

The unimodality of the halo VVFs is a feature shared by all the tracer samples we consider in this work, and allows us to equivalently describe each VVF by simply reporting a small number of its percentiles. As an example, we have indicated the percentiles $y_{50}$ (solid), $y_{16},y_{84}$ (dashed) and $y_{2.5},y_{97.5}$ (dotted) of the Poisson VVF as vertical lines, using which one can read off the median, central $68\%$ and $95\%$ regions, respectively, of the distribution. (Hereafter, we will refer to the $p^{\rm th}$ percentile of $y$ as $y_p$.) In subsequent plots, we will exclusively display VVF percentiles instead of differential distributions for all tracer samples, comparing with the Poisson VVF percentiles. This will allow us to compactly represent the VVF shapes of multiple tracer populations on the same graph.

\begin{figure}
\centering
\includegraphics[width=0.45\textwidth]{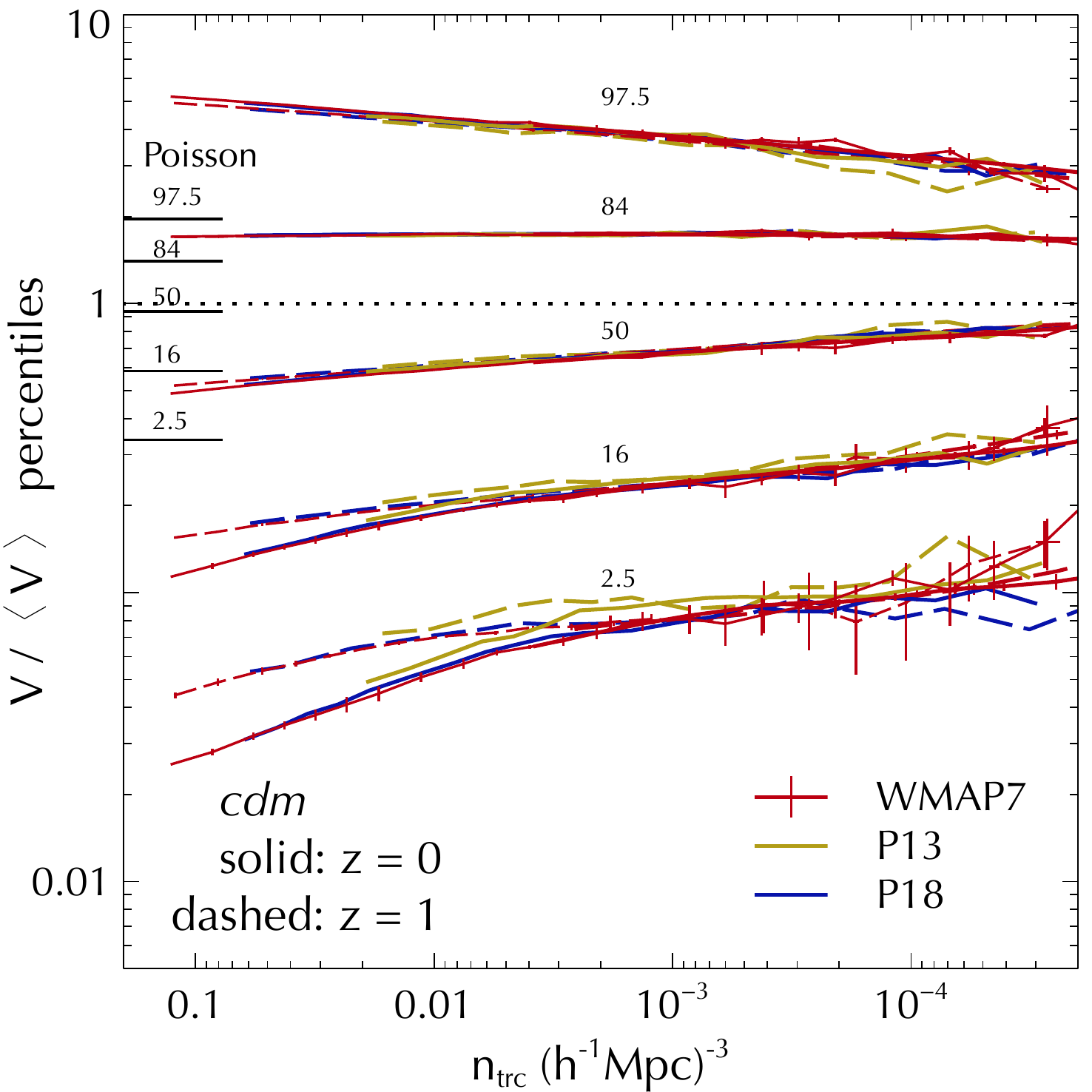}
\caption{{\bf Percentiles of the VVF} for real-space haloes selected by a mass threshold $m_{\rm 200b}>m_{\rm lim}$ at $z=0$ (solid) and $z=1$ (dashed) for three CDM cosmologies: WMAP7 (red), P13 (yellow) and P18 (blue). We display the percentiles $y_{2.5}$, $y_{16}$, $y_{50}$, $y_{84}$ and $y_{97.5}$ (from bottom to top, as labelled) as a function of tracer number density $n_{\rm trc}$ (see Figure~\ref{fig:hmf} for the corresponding $m_{\rm lim}$ values). For comparison, the VVF percentiles of Poisson distributed tracers computed using \eqn{eq:Poisson-p(y)} are indicated as horizontal line segments at the left of the plot. 
WMAP7 results are averaged over all available realisations of the configurations ${\rm L}150\_{\rm N}1024$ and ${\rm L}600\_{\rm N}1024$ (see Table~\ref{tab:sims}) with error bars indicating the standard deviation across realisations.
Horizontal dotted line indicates the mean value $\avg{y}=1$.
}
\label{fig:vvfperc-ntrc}
\end{figure}

Figure~\ref{fig:vvfperc-ntrc} shows the VVF percentiles of various mass-thresholded samples as a function of their tracer number density $n_{\rm trc}(>m_{\rm lim})$ (see Figure~\ref{fig:hmf}). 
Note that, for each value of mass threshold $m_{\rm lim}$, we use all selected haloes to perform the tessellation and results are then shown for different mass thresholds using $n_{\rm trc}(>m_{\rm lim})$ rather than $m_{\rm lim}$ as the control variable.
We display the percentiles $y_{2.5},y_{16},y_{50},y_{84}$ and $y_{97.5}$, showing results at $z=0$ (solid lines) and $z=1$ (dashed lines).

As noted earlier, all distributions are broader than the Poisson case (computed using equation~\ref{eq:Poisson-p(y)} and shown as horizontal line segments at the left of the plot). Comparing between the different CDM cosmologies, we see essentially universal behaviour for all the percentiles at fixed redshift.
Additionally, there is a substantial redshift evolution of the lower percentiles $y_{16}$ and especially $y_{2.5}$ (corresponding to tracers in high density regions), with much milder evolution in the upper percentiles. 

\begin{figure}
\centering
\includegraphics[width=0.45\textwidth]{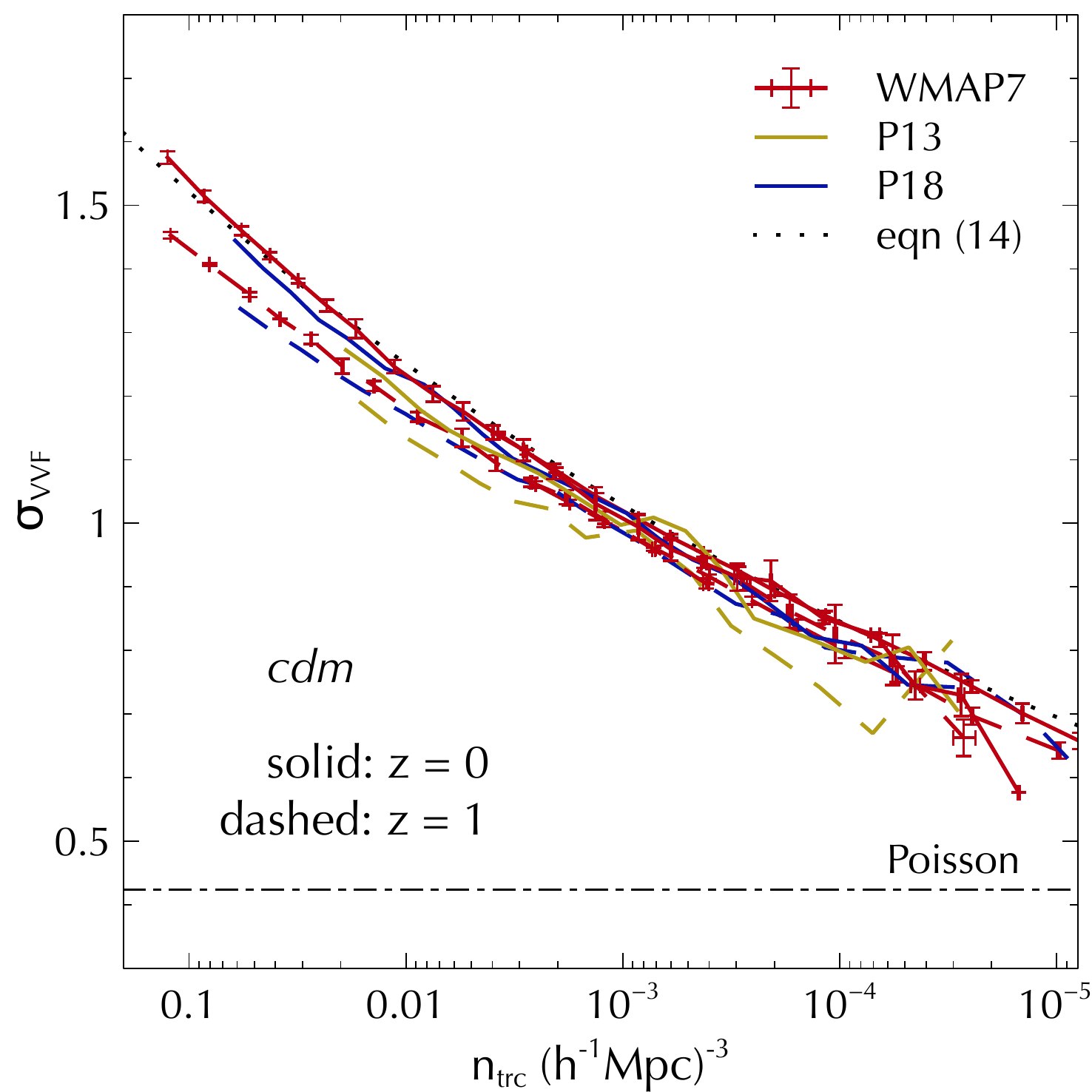}
\caption{{\bf Standard deviation of the VVF} ($\sigma_{\rm VVF}$; equation~\ref{eq:stdVVF}) for the same halo samples used in Figure~\ref{fig:vvfperc-ntrc} and formatted identically. The thin dotted curve shows the fit in \eqn{eq:stdVVF-fit-ntrc} while the horizontal dash-dotted line indicates the constant value of $\sigma_{\rm VVF}$ for Poisson distributed tracers.
WMAP7 results are averaged over all available realisations of the configurations ${\rm L}150\_{\rm N}1024$ and ${\rm L}600\_{\rm N}1024$ (see Table~\ref{tab:sims}) with error bars indicating the standard deviation across realisations.}
\label{fig:vvfstd-ntrc}
\end{figure}

A compact description of the width of the VVF is provided by the standard deviation $\sigma_{\rm VVF}$ given by
\be
\sigma_{\rm VVF} = \sqrt{\avg{y^2}-1}\,,
\label{eq:stdVVF}
\ee
which is also formally easier to describe than the full VVF, as we saw in section~\ref{subsec:theory:clustered} (see equation~\ref{eq:<y^2>}). 
Figure~\ref{fig:vvfstd-ntrc} shows the measured values of $\sigma_{\rm VVF}$ for the same mass-thresholded samples discussed above. 
Consistently with the behaviour of the percentiles of the VVF, we see that $\sigma_{\rm VVF}$ is a nearly universal function of $n_{\rm trc}$ regardless of cosmology, with only mild redshift evolution. 
The dotted curve shows a `by-eye' fit to the $z=0$ WMAP7 result (which describes all the CDM results at better than $\sim5\%$) given by
\be
\sigma_{\rm VVF}(n_{\rm trc}) = 1.85 \times \left(\frac{n_{\rm trc}}{1(\Mpch)^{-3}}\right)^{0.085}\,.
\label{eq:stdVVF-fit-ntrc}
\ee

\begin{figure*}
\centering
\includegraphics[width=1.0\textwidth]{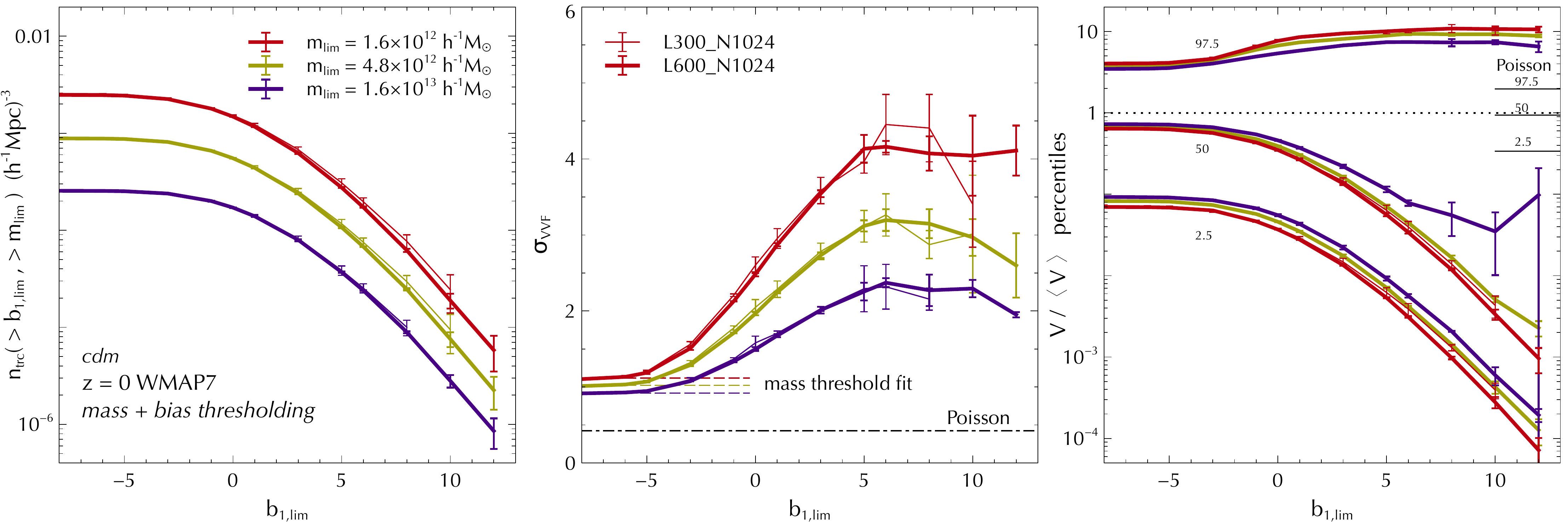}
\caption{{\bf Effects of halo clustering:} 
\emph{(Left panel:)} Halo number density as a function of linear bias threshold $b_{\rm 1,lim}$, for three different mass thresholds $m_{\rm 200b}>m_{\rm lim}$.  
\emph{(Middle panel:)} VVF standard deviation $\sigma_{\rm VVF}$ for the corresponding samples. The horizontal dashed line segments show the asymptotic values expected from combining the fit in \eqn{eq:stdVVF-fit-ntrc} with the T08 mass function for the corresponding thresholds. Horizontal dash-dotted line indicates the value for Poisson distributed tracers.
\emph{(Right panel:)} VVF percentiles for the corresponding samples. Horizontal line segments indicate the Poisson values.
Results are displayed for parent halo samples at $z=0$ in the WMAP7 CDM configurations. Curves of different thickness correspond to different box configurations as indicated in the legend of the \emph{middle panel}.
All results are averaged over all available realisations (see Table~\ref{tab:sims}) with error bars indicating the standard deviation across realisations. The VVF shape is clearly a strong function of the $b_1$ threshold; see text for a discussion.}
\label{fig:vvfntrcstdperc-bias}
\end{figure*}

\subsection{Effects beyond halo mass}
\label{subsec:beyondmass}
\noindent
Although the VVF of real-space haloes selected by mass is evidently a near-universal function of number density,
the presence of the infinite hierarchy of tracer correlation functions in its construction makes it interesting to investigate the role played by various aspects of halo clustering in determining the shape of the VVF.  
 In the following, we will explore the effects of large-scale linear halo bias, the presence of substructure and redshift space distortions (RSD). Additionally, in Appendix~\ref{app:downmaskassembly} we discuss the role of downsampling (relevant for mapping dark haloes to galaxies), masking (relevant for modelling observed samples) and assembly bias. 
For this part of the analysis, we display results only for the WMAP7 simulations at $z=0$. 

\subsubsection{Large-scale linear bias}
\label{subsubsec:b1}
\noindent
In order to understand the role of halo clustering, it is convenient to use the large-scale halo-by-halo linear bias estimate introduced by \citet{phs18a}. This is essentially a Fourier space calculation of the ratio of the halo-matter cross-power spectrum $P_\times(k)$ to the matter power spectrum $P(k)$, except that $P_\times(k)$ is calculated for one halo at a time. The linear bias $b_1$ for each halo is then estimated as a weighted sum of the ratio $P_\times(k)/P(k)$ over low-$k$ modes.\footnote{We introduce one modification to this calculation as compared to \citet{phs18a}. Those authors used weights proportional to the number of modes $N_k$ in each  $k$-bin to calculate $b_1\sim\sum_kN_kP_\times(k)/P(k) / \sum_kN_k$, which is the least squares estimator under the assumption of Gaussian errors when the number of haloes in the cross-power calculation is large. Since we are treating one halo at a time, there are additional terms which must be included in calculating the noise of the power spectrum estimator \citep{smith09}; we show in Appendix~\ref{app:hbyhbias} that the appropriate weights are then proportional to $N_kP(k)$ to get the least squares estimator $b_1\sim\sum_kN_kP_\times(k) / \sum_kN_kP(k)$.}

To explore the effects of halo clustering on the VVF, we construct halo samples by first imposing a mass threshold $m_{\rm 200b} > m_{\rm lim}$ and then further imposing a bias threshold $b_1>b_{\rm 1,lim}$. The \emph{left panel} of Figure~\ref{fig:vvfntrcstdperc-bias} shows the tracer number density for such samples as a function of $b_{\rm 1,lim}$ for three choices of $m_{\rm lim}$. The \emph{middle panel} of the Figure shows the corresponding values of $\sigma_{\rm VVF}$. Compared to the purely mass-thresholded values which are achieved asymptotically at the lowest $b_{\rm 1,lim}$ (indicated by the dashed line segments), we see that $\sigma_{\rm VVF}$ is a strong function of $b_{\rm 1,lim}$ (note the range on the vertical axis). Moreover, the span of values of $\sigma_{\rm VVF}$ across the three $m_{\rm lim}$ values also increases significantly at higher $b_{\rm 1,lim}$.

The \emph{middle panel} of Figure~\ref{fig:vvfntrcstdperc-bias} also shows another interesting feature: although $\sigma_{\rm VVF}$ decreases with increasing $m_{\rm lim}$ for any fixed $b_{\rm 1,lim}$, it shows more complex behaviour with increasing $b_{\rm 1,lim}$ at fixed $m_{\rm lim}$, first increasing and then flattening (or even decreasing, although this may be related to finite volume effects) for $b_{\rm 1,lim}\gtrsim6$. This clearly indicates that halo mass and large-scale clustering are two independent variables determining the VVF; the naive expectation that `high mass equals high bias' does not work for the VVF. This exemplifies the presence of the infinite hierarchy of $N$-point correlation functions in the VVF, with $b_1$ representing the effects of large-scale clustering and halo mass of small-scale non-linearities. 

Finally, the \emph{right panel} of Figure~\ref{fig:vvfntrcstdperc-bias} shows the VVF percentiles for these mass+bias selected halo samples. We see that, unlike the purely mass-thresholded samples studied earlier, in this case the \emph{median} value of $y$ is a very steeply decreasing function of $b_{\rm 1,lim}$ for any $m_{\rm lim}$, while the upper-most percentile increases with $b_{\rm 1,lim}$. This is qualitatively different from the behaviour as a function of $m_{\rm lim}$ or its corresponding $n_{\rm trc}$ seen, e.g., in Figure~\ref{fig:vvfperc-ntrc}.

\subsubsection{Substructure}
\label{subsubsec:sub}
\noindent
So far we have been dealing with parent haloes, which are expected to host central galaxies. Galaxy catalogs typically also contain a substantial fraction of satellite galaxies which occupy subhaloes of larger systems. Compared to the Voronoi cell structure in a catalog containing only centrals/parent haloes, a catalog containing satellites/subhaloes would contain preferentially smaller Voronoi cells, since the inclusion of satellites in a group would split the erstwhile Voronoi cell of the group's central into smaller chunks. This effect would be more pronounced at smaller thresholds $m_{\rm lim}$ where the substructure fraction is higher. We study this effect here using subhaloes in our $N$-body simulations.

\begin{figure*}
\centering
\includegraphics[width=0.85\textwidth]{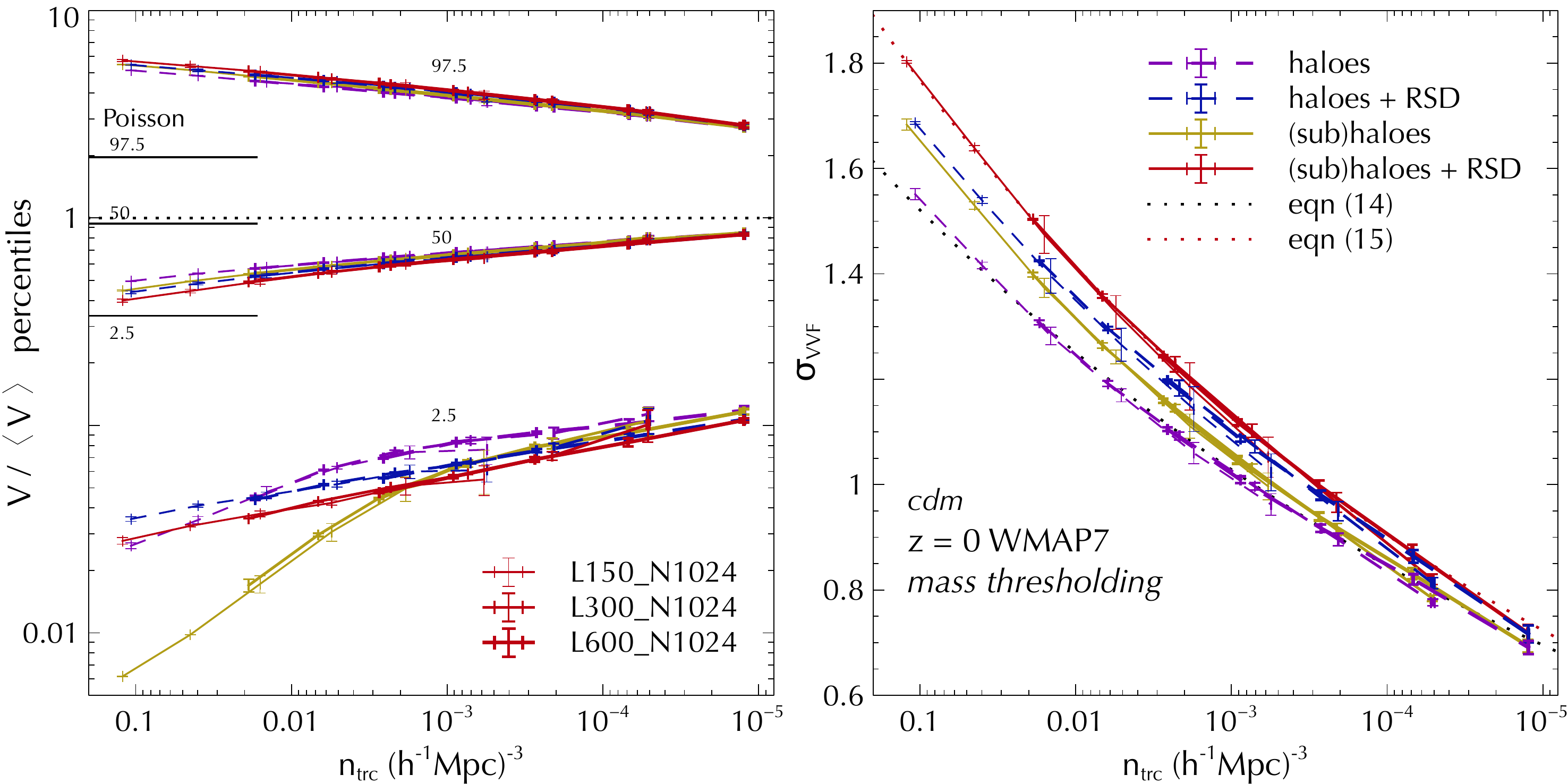}
\caption{{\bf Comparing effects of RSD and substructure:} 
VVF percentiles \emph{(left panel)} and standard deviation $\sigma_{\rm VVF}$ \emph{(right panel)} for mass-thresholded samples containing only parent haloes (dashed) in real space (purple) and redshift space (blue), and for samples additionally containing subhaloes (solid) in real (yellow) and redshift space (red). 
Dotted curves in the \emph{right panel} show the fits from \eqn{eq:stdVVF-fit-ntrc} (black, lower) and \eqn{eq:stdVVF-fit-ntrc-RSDsub} (red, upper).
Results are displayed for tracers at $z=0$ in the WMAP7 CDM configurations. Curves of different thickness correspond to different box configurations as indicated in the legend of the \emph{left panel}.
All results are averaged over all available realisations (see Table~\ref{tab:sims}) with error bars indicating the standard deviation across realisations. 
}
\label{fig:vvfpercstd-RSDsub-CDM}
\end{figure*}

The yellow solid curves in Figure~\ref{fig:vvfpercstd-RSDsub-CDM} show the VVF percentiles \emph{(left panel)} and standard deviation \emph{(right panel)} for samples containing all haloes and subhaloes with $m_{\rm 200b}>m_{\rm lim}$\footnote{Strictly speaking, one should account for the effects of tidal stripping by thresholding on mass definitions such as $m_{\rm peak}$ which would account for the subhalo's entire accretion history. We will do so later when comparing our simulations with observational results. For now, we stick to the $m_{\rm 200b}$ mass definition which allows us to use our full suite of simulation configurations; this would otherwise be curtailed due to the absence of merger trees for boxes with $L_{\rm box}\geq300\Mpch$.} as a function of the corresponding $n_{\rm trc}$. The purple dashed curves show the corresponding measurements for samples containing only parent haloes (repeated from Figures~\ref{fig:vvfperc-ntrc} and~\ref{fig:vvfstd-ntrc}). 
We see that the VVF distribution -- particularly at small $m_{\rm lim}$ (large $n_{\rm trc}$) -- broadens towards smaller values of $y$ and has a larger width $\sigma_{\rm VVF}$ upon including subhaloes. The percentile $y_{2.5}$ has a pronounced knee-like feature around $n_{\rm trc}\sim10^{-3}(\Mpch)^{-3}$. Thus, the additional clustering information introduced by substructure produces large effects in the small-volume (or high-density) tail of the VVF. This can be potentially very interesting for studies of galaxy groups, e.g., by placing constraints on the outputs of group-finder algorithms. It is, however, important to first assess the role of RSD which can substantially alter the observed spatial distribution of substructure due to line-of-sight virial motions. We turn to this next.

\subsubsection{Redshift space distortions}
\label{subsubsec:RSD}
\noindent
The blue dashed curves in Figure~\ref{fig:vvfpercstd-RSDsub-CDM} show the VVF statistics for the same parent haloes used for the purple dashed curves, but first moved into redshift space under the distant observer approximation by choosing one of the simulation box axes as the observer line-of-sight.  Since this is a parent-only sample, virial motions are expected to play no role and the entire RSD effect should be due to large-scale bulk flows (except possibly when there is a contamination of the sample by splashback objects, see below). We see that $\sigma_{\rm VVF}$ for the redshift-space halo sample is always systematically larger than its real-space counterpart. Correspondingly, $y_{2.5}$ for the redshift-space sample is lower than the real-space one, at all but the largest number densities. Since large-scale bulk flows enhance the large-scale bias \citep{kaiser87}, and we have already seen that $\sigma_{\rm VVF}$ is a strong function of halo bias (c.f. Figure~\ref{fig:vvfntrcstdperc-bias}), the enhancement of $\sigma_{\rm VVF}$ in redshift space compared to real space is not surprising. This is also consistent with $y_{2.5}$ in redshift space being lower than that in real space for most of the samples. To understand the reversal of the latter trend in the high-$n_{\rm trc}$ samples ($n_{\rm trc}\gtrsim10^{-2}(\Mpch)^{-3}$), it is useful to first consider the effect of substructure.

We next include subhaloes in the samples as in section~\ref{subsubsec:sub} and move all objects into redshift space. Our samples are now affected not only by bulk flows but also by the Fingers-of-God effect due to virial motions of subhaloes in groups. The red solid curves in Figure~\ref{fig:vvfpercstd-RSDsub-CDM} show the resulting VVF statistics. The standard deviation $\sigma_{\rm VVF}$ in the \emph{right panel} is systematically enhanced compared to all other samples, a sign of the doubly enhanced clustering due to both bulk flows and the presence of substructure. 

\begin{figure*}
\centering
\includegraphics[width=0.85\textwidth]{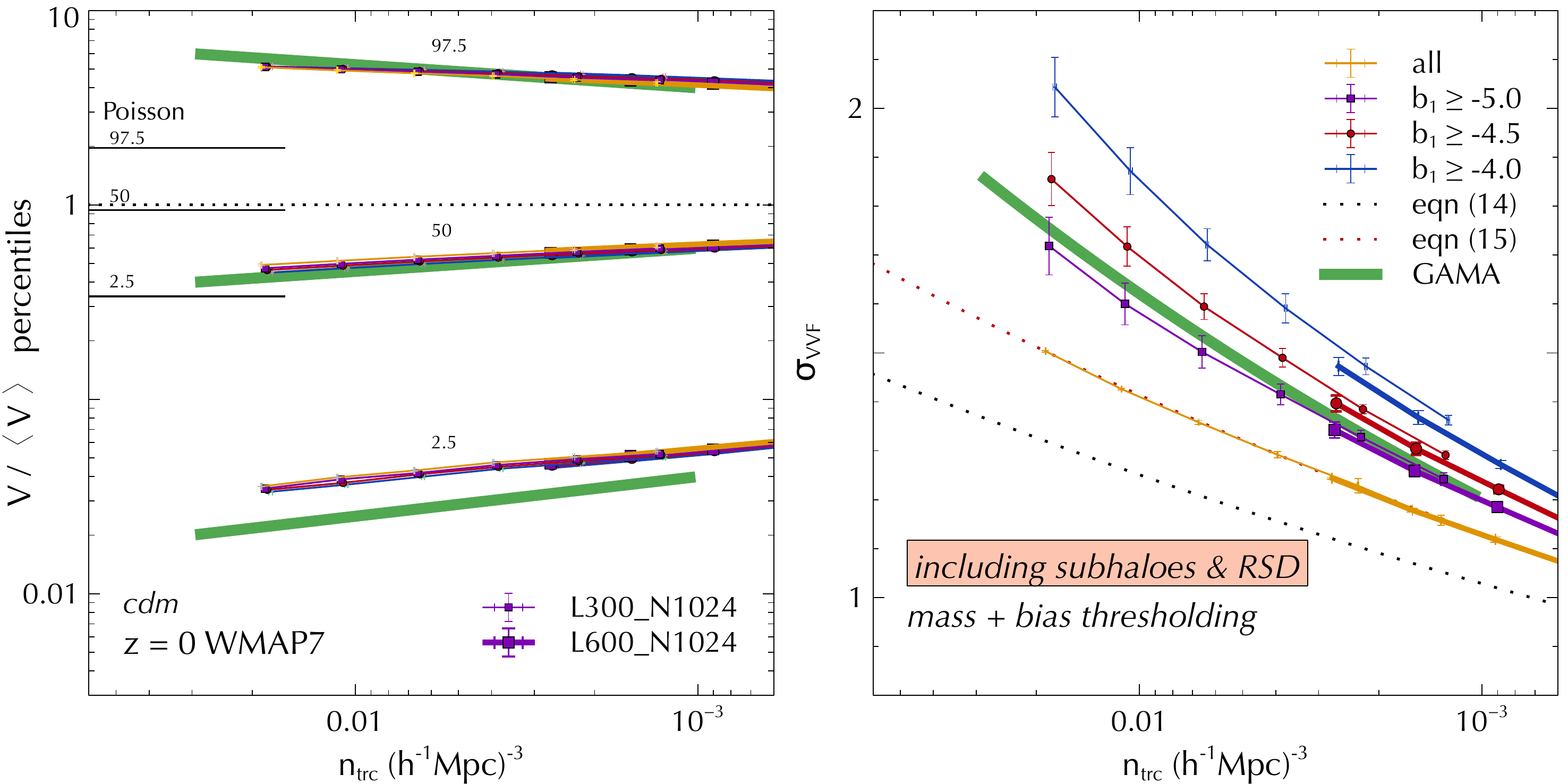}
\caption{{\bf Sensitivity to halo bias:} VVF percentiles \emph{(left panel)} and standard deviation $\sigma_{\rm VVF}$ \emph{(right panel)} for redshift-space samples including subhaloes (solid curves) selected by a threshold on mass $m_{\rm 200b}$ (yellow) and additionally imposing a threshold on bias $b_1$ (purple, red and blue, as indicated).
Dotted curves in the \emph{right panel} are repeated from Figure~\ref{fig:vvfpercstd-RSDsub-CDM} and show the fits from \eqns{eq:stdVVF-fit-ntrc} (black) and~\eqref{eq:stdVVF-fit-ntrc-RSDsub} (red).
Results are displayed for the $z=0$ WMAP7 CDM simulations for two configurations as indicated in the legend of the \emph{left panel}. Results were averaged over all available realisations (see Table~\ref{tab:sims}) with error bars indicating the standard deviation across realisations.
Thick solid green curves show the power law fits to GAMA measurements from Table~\ref{tab:GAMA}. The range of these curves in $n_{\rm trc}$ corresponds to $r$-band absolute magnitude thresholds between $-18$ and $-21$.
GAMA results for $\sigma_{\rm VVF}$ in the \emph{right panel} are bracketed between the bias thresholds $b_1\geq-5$ and $b_1\geq-4.5$. None of the samples, however, can describe the GAMA results for the percentile $y_{2.5}$ in the \emph{left panel}. See text for a discussion.}
\label{fig:vvfpercstd-bias-RSDsub}
\end{figure*}

More interestingly, $y_{2.5}$ in the \emph{left panel} now shows a dramatic difference as compared to the real-space sample with substructure: the knee-like feature has completely disappeared and the final result is close to being a single power-law in $n_{\rm trc}$. The enhancement is easily understood as being due to the preferential elongation of all groups along the observer line-of-sight which stretches out all subhalo Voronoi cells along this direction and increases their volumes. Comparing the solid red and dashed blue curves for $y_{2.5}$, the RSD effect for the parent-only sample visually appears to be simply a milder version of the drastic flattening seen in the sample containing subhaloes. 
This could either be caused by backsplash objects -- subhaloes mimicking isolated objects by being located temporarily far from their host \citep{gkg05} -- contaminating the parent-only samples, or more generally due to preferential flows in regions with strong tidal anisotropy, such as near the nodes of thick filaments.
In either case, the lower percentiles of the VVF are clearly sensitive to such dynamical effects. 
This could potentially be of great practical value in both theoretical and observational studies that, e.g., seek to robustly separate central objects from substructure, or characterise the dynamics within different cosmic web environments. 

Overall, upon including subhaloes as well as RSD in the otherwise mass-thresholded samples, we see that $\sigma_{\rm VVF}$ as well as the percentiles of $y$ become nearly single power-laws in $n_{\rm trc}$. We find that $\sigma_{\rm VVF}(n_{\rm trc})$ is now well-described by 
\be
\sigma_{\rm VVF|RSD+sub}(n_{\rm trc}) = 2.21 \times \left(\frac{n_{\rm trc}}{1(\Mpch)^{-3}}\right)^{0.097}\,,
\label{eq:stdVVF-fit-ntrc-RSDsub}
\ee
(shown as the red dotted curve in the right panel of Figure~\ref{fig:vvfpercstd-RSDsub-CDM}). 

Appendix~\ref{app:downmaskassembly} further shows that the effects of downsampling lead to a characteristic decrease in the width of the VVF, whose effect on $\sigma_{\rm VVF}$ is accurately captured by the separable form in \eqn{eq:stdVVF-fit-ntrcfd} for both parent haloes in real-space as well as redshift-space samples including substructure, while masking leaves no discernable imprint on the VVF. With this understanding of the dependence of the VVF on variables related to halo clustering, we next turn to a comparison with observed galaxy samples.

\section{Matching GAMA Results}
\label{sec:GAMA}
\noindent
In a forthcoming paper (Alam et al., in preparation; henceforth, Paper-II), we analyse luminosity-thresholded samples in the  Galaxies \& Mass Assembly (GAMA) survey \citep{driver+09},\footnote{\href{http://www.gama-survey.org}{http://www.gama-survey.org}} constructing the Voronoi tessellation for each sample and measuring the corresponding VVF. The GAMA survey comprises a spectroscopic sample of $\sim300,000$ galaxies with a magnitude limit $r<19.8$ in an area of $\sim286{\rm deg}^2$ with approximately $98\%$ completeness. Details of our analysis can be found Paper-II, where we show that the GAMA VVF is well-described by simple power law relations for $\sigma_{\rm VVF}$ and the percentiles $y_p$, $p\in\{2.5,50,97.5\}$, as a function of $n_{\rm trc}$ for samples thresholded by $r$-band absolute magnitude, with threshold values ranging from $-21$ to $-18$. Table~\ref{tab:GAMA} summarises these power laws for the publicly available G15 sample. 
In this section, we construct a halo sample that best describes the VVF of the GAMA G15 field. 

\begin{table}
\centering
\begin{tabular}{ccc}
\hline
\hline
\textrm{Statistic} & $A$ & $\alpha$\\
\hline
\hline
$\sigma_{\rm VVF}$ & 2.95 & 0.130\\
$y_{2.5}$ & $4.84\times10^{-3}$ & $-0.306$\\
$y_{50}$ & $0.263$ & $-0.119$\\
$y_{97.5}$ & $9.1$ & $0.119$\\
\hline
\end{tabular}
\caption{Power law descriptions $A\times n_{\rm trc}^\alpha$ -- with $n_{\rm trc}$ in units of $(\Mpch)^{-3}$ -- of GAMA measurements of the standard deviation $\sigma_{\rm VVF}$ and percentiles $y_{2.5}$, $y_{50}$ and $y_{97.5}$ of the VVF of luminosity-thresholded samples. These descriptions are valid over the range $10^{-3}\lesssim n_{\rm trc}\lesssim3\times10^{-2}$, corresponding to $r$-band absolute magnitude thresholds between $-21$ and $-18$.}
\label{tab:GAMA}
\end{table}

Figures~\ref{fig:vvfpercstd-down-RSDsub} and~\ref{fig:vvfpercstd-mask-RSDsub} show that, when samples are thresholded by $m_{\rm 200b}$, the GAMA results cannot be explained by the inclusion of subhaloes in the sample, RSD, downsampling or masking. This leaves variations in large-scale halo bias as a likely potential explanation. In Figure~\ref{fig:vvfpercstd-bias-RSDsub} we investigate the sensitivity of the VVF of mass-thresholded samples, which include subhaloes and RSD, to small changes in $b_1$. To focus the discussion, we restrict attention to the two statistics $\sigma_{\rm VVF}$ and $y_{2.5}$. The all-(sub)halo $m_{\rm 200b}$-selected sample does not describe either of these statistics well. Upon changing the halo bias by imposing successively larger bias thresholds, however, we indeed see that $\sigma_{\rm VVF}$ is very nicely bracketed between $b_1$ thresholds of $-5$ and $-4.5$. 

The percentile $y_{2.5}$, on the other hand, is very insensitive to the bias threshold and is consequently not well-described by any of the samples thresholded by $m_{\rm 200b}$ and $b_1$.
Since $y_{2.5}$, particularly at high number densities, is expected to be sensitive to the behaviour of the subhalo population (c.f. Figure~\ref{fig:vvfpercstd-RSDsub-CDM}), this suggests that our choice of $m_{\rm 200b}$-thresholding is not picking the correct population of subhaloes. This is not surprising: subhaloes are dramatically affected by tidal stripping \citep[e.g.,][]{vdbg18} and might easily fail a cut on $m_{\rm 200b}$ while still being valid candidates for hosting faint galaxies. 

\begin{figure*}
\centering
\includegraphics[width=0.95\textwidth]{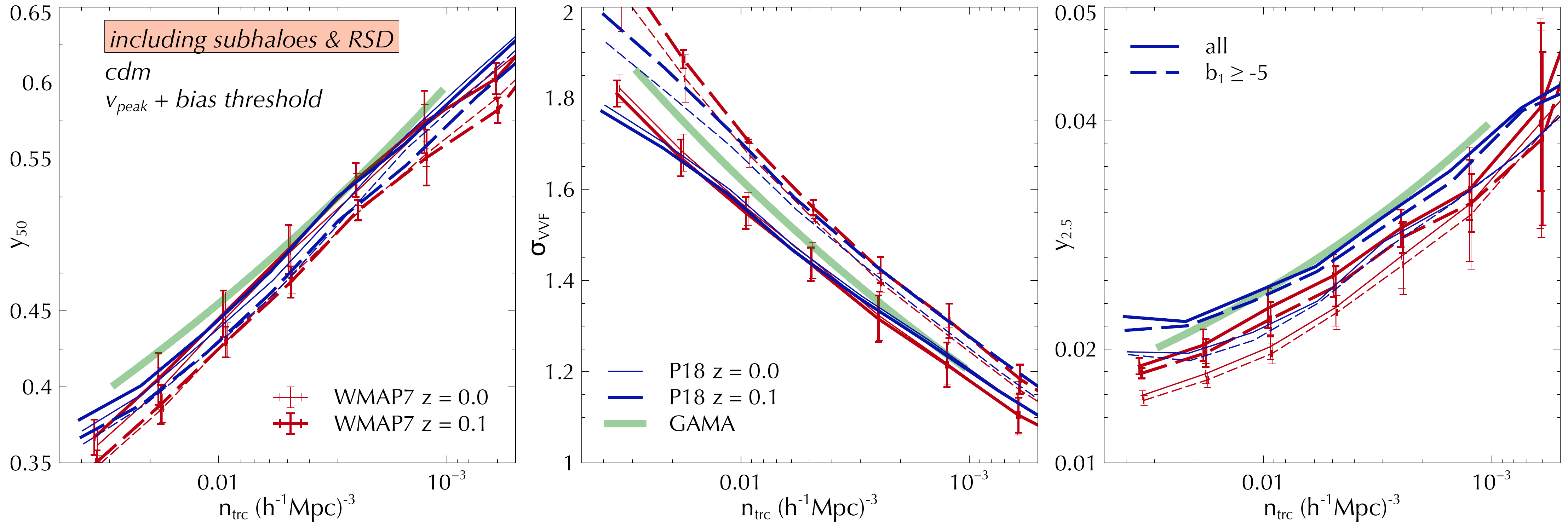}
\caption{{\bf GAMA-like sample (SHAM with $v_{\rm peak}$):} 
VVF median \emph{(left panel)}, standard deviation $\sigma_{\rm VVF}$ \emph{(middle panel)} and percentile $y_{2.5}$ \emph{(right panel)} for redshift-space samples including subhaloes selected by a threshold on $v_{\rm peak}$ (solid) and additionally imposing a bias threshold $b_1 \geq -5$ (dashed). 
The thick (thin) curves show results at redshift $z=0.1$ ($z=0.0$) for the P18 (blue) and WMAP7 (red) CDM cosmologies.
WMAP7 results were averaged over 2 realisations of the ${\rm L}150\_{\rm N}1024$ configuration, with error bars showing the standard deviation across the realisations.
Thick solid green curves show the power law fits to GAMA measurements from Table~\ref{tab:GAMA}.
The GAMA median and $y_{2.5}$ are best described by the P18 results at $z=0.1$, regardless of selecting on $b_1$ (although the median seems to prefer the full sample). The GAMA $\sigma_{\rm VVF}$, on the other hand, is bracketed between the full sample and the bias-thresholded one for either cosmology, roughly independent of redshift.
See text for a discussion.}
\label{fig:vvf-vpeak-GAMA}
\end{figure*}

To rectify the exclusion of such objects, in Figure~\ref{fig:vvf-vpeak-GAMA} we show results for samples thresholded by $v_{\rm peak}$, which is the maximum value of the maximum circular velocity of an object along its main progenitor branch of the merger tree. Subhalo abundance matching (SHAM) linking $v_{\rm peak}$ with stellar mass $m_\ast$ is known to provide a good description of the $m_\ast$-dependent 2-point clustering of low-redshift galaxies with $m_\ast\gtrsim10^{9.8}\Mhsq$ (approximately corresponding to $r$-band absolute magnitude $M_r-5\log_{10}h\lesssim-19$) at all but the smallest length scales \citep{rwtb13,campbell+18}.\footnote{For simplicity we assume that the $v_{\rm peak}$-luminosity SHAM implied by our comparison below is equivalent to the $v_{\rm peak}$-$m_\ast$ SHAM usually studied in the literature \citep[see also][]{gerke+13,carretero+15}. We will test this assumption in future work.}
Figure~\ref{fig:vvf-vpeak-GAMA} shows results for the P18 (blue) and WMAP7 (red) CDM simulations, for samples selected with (dashed) and without (solid) a bias threshold,\footnote{We caution that the bias-thresholded WMAP7 results are likely affected by finite volume effects. We have checked that a similar analysis with $m_{\rm 200b}$-thresholding (c.f. Figure~\ref{fig:vvf-vpeak-GAMA}) leads to $\sigma_{\rm VVF}$ being systematically underestimated by $\sim5\%$ in the WMAP7 CDM ${\rm L}150\_{\rm N}1024$ boxes.} and also explores the redshift evolution of the VVF statistics by comparing results at $z=0$ (thin lines) with those at $z=0.1$ (thick lines) which is closer to the median redshift of GAMA galaxies.

We see that the percentile $y_{2.5}$ of GAMA \emph{(right panel)} is best described by the P18 sample at $z=0.1$, regardless of the $b_1$ threshold, while P18 results at $z=0.0$ and all the WMAP7 results seem to be excluded. The GAMA median $y_{50}$ \emph{(left panel)}, on the other hand, seems to prefer the full samples in either cosmology compared to their $b_1\geq-5$ counterparts, independent of redshift. Finally, the GAMA $\sigma_{\rm VVF}$ \emph{(middle panel)} is bracketed between the all (sub)halo and $b_1\geq-5$ samples for both cosmologies at each redshift. 

Thus, different aspects of the VVF allow us to simultaneously probe the effects of galaxy evolution (as captured by the environment-dependence implied by the $b_1$ selection) and cosmology. The median $y_{50}$ and standard deviation $\sigma_{\rm VVF}$ are pulled in different directions by the $b_1$ threshold, while $y_{2.5}$ is more sensitive to cosmology, redshift evolution and the nature of substructure (c.f. Figures~\ref{fig:vvfpercstd-RSDsub-CDM} and~\ref{fig:vvfpercstd-bias-RSDsub}). Finally, since SHAM with $v_{\rm peak}$ (with no restrictions on large-scale environment) reproduces the observed 2-point correlation function of low-redshift galaxies, \emph{the sensitivity of $y_{50}$ and $\sigma_{\rm VVF}$ to halo bias is genuinely new information which is not easily accessible to traditional analyses.} This is of great interest for studies aimed at detecting beyond-mass effects in galaxy evolution \citep[see, e.g.,][who claimed evidence of beyond-mass effects, using observational proxies of halo formation time, as a function of cosmic web environment in the GAMA survey]{tojeiro+17}.

We emphasise that our discussion above has completely ignored errors on the GAMA measurements; consequently, our conclusions regarding the acceptability of different models are indicative only. In Paper-II, we will perform a rigorous parameter inference study using the GAMA VVF measurements. In the next section, we further explore the dependence of these observables on the adopted cosmological model.

\section{Cosmology dependence}
\label{sec:cosmo}
\noindent
We end our analysis here by showing the VVF results for the $v_{\rm peak}$ SHAM at $z=0.1$ for a sampling of cosmological models. These include the WMAP7 and P18 CDM models discussed so far in the text, as well as two non-standard dark matter models which we describe next. 

The first of these is a warm dark matter (WDM) model in which the matter power spectrum at early epochs is suppressed at small scales due to the free-streaming of a thermally produced WDM particle with mass $m_{\rm dm}=0.4{\rm keV}$. Although such a particle is completely ruled out by Lyman-alpha forest observations as being the dominant component of dark matter \citep{Viel2013,irsic+17,pd+20,grmb20}, it allows us to resolve the entire initial power spectrum up to the truncation scale with sufficient particles and thus serves as a useful extreme toy model for our investigation of signatures in the VVF. In particular, we modify the CDM transfer function according to the fitting function of \citet{Bode2001} \citep[with parameters taken from][]{Viel2005} 
\begin{equation}
 T_{\rm wdm}(k) = T_{\rm cdm}(k) \left[1 + (\alpha k)^{2\mu}\right]^{-5/\mu},
\label{eq:Tk-wdm}
\end{equation}
with $\mu=1.12$ and
\begin{equation}
\alpha  \equiv 0.049 \left(\frac{\Omega_{\rm m}}{0.25}\right)^{0.11} \left(\frac{h}{0.7}\right)^{1.22}
\left( \frac{m_{\rm dm}}{1\,{\rm keV}}  \right)^{-1.11}\,\Mpch\,.
\label{alpha}
\end{equation}
This results in a ``half-mode'' mass-scale \citep[c.f., e.g.,][]{Schneider2012} of $M_{\rm hm}\simeq 10^{12}\Mh$, which is resolved with $\sim4200$ particles by our ${\rm L}150\_{\rm N}1024$ WMAP7 configuration.\footnote{Our simulations treat the collisionless WDM fluid in the perfectly cold limit, ignoring the (small) thermal velocity dispersion that a real WDM fluid would possess. This is expected to be accurate at late times, well after perturbations have been suppressed below the maximum free-streaming scale in linear perturbation theory \citep{Angulo2013}.} We generated the initial conditions for this simulation using the same random seed as for one of the two ${\rm L}150\_{\rm N}1024$ WMAP7 CDM realisations.

The second non-standard model we explore displays a different type of small-scale feature -- namely, oscillations -- in the initial power spectrum. Such models have been discussed extensively in the context of possible acoustic oscillations due to interactions in the dark sector \citep{cyr-racine+16,vogelsberger+16}. We focus on
the so-called ``ballistic'' dark matter BDM model recently proposed by \citet{ddk19}. 

This model contains a dark matter component which remained relativistic and collisional until relatively late times, before becoming cold and collisionless through a phase transition. The acoustic oscillations of this component prior to this phase transition leave an imprint during its collisionless phase in the form of large coherent peculiar velocities which leads to a ballistic evolution (hence the name) until these velocities are damped away due to Hubble expansion. As a result, the initial power spectra for non-linear structure formation in such a model contain oscillatory features whose peak structure and overall amplitude is sensitive to the fraction $f_{\rm bdm}$ of the total dark matter that comprises the ballistic species and the redshift $z_\ast$ of the phase transition. Here we choose $z_\ast=10^5$ and $f_{\rm bdm}=0.5$ which are approximately consistent with \emph{Planck} measurements of CMB anisotropies at the $\sim2\sigma$ level \citep{ddk19}. The initial conditions for this simulation were generated using the same random seed as for the ${\rm L}200\_{\rm N}1024$ P18 CDM simulation.

\begin{figure}
\centering
\includegraphics[width=0.45\textwidth]{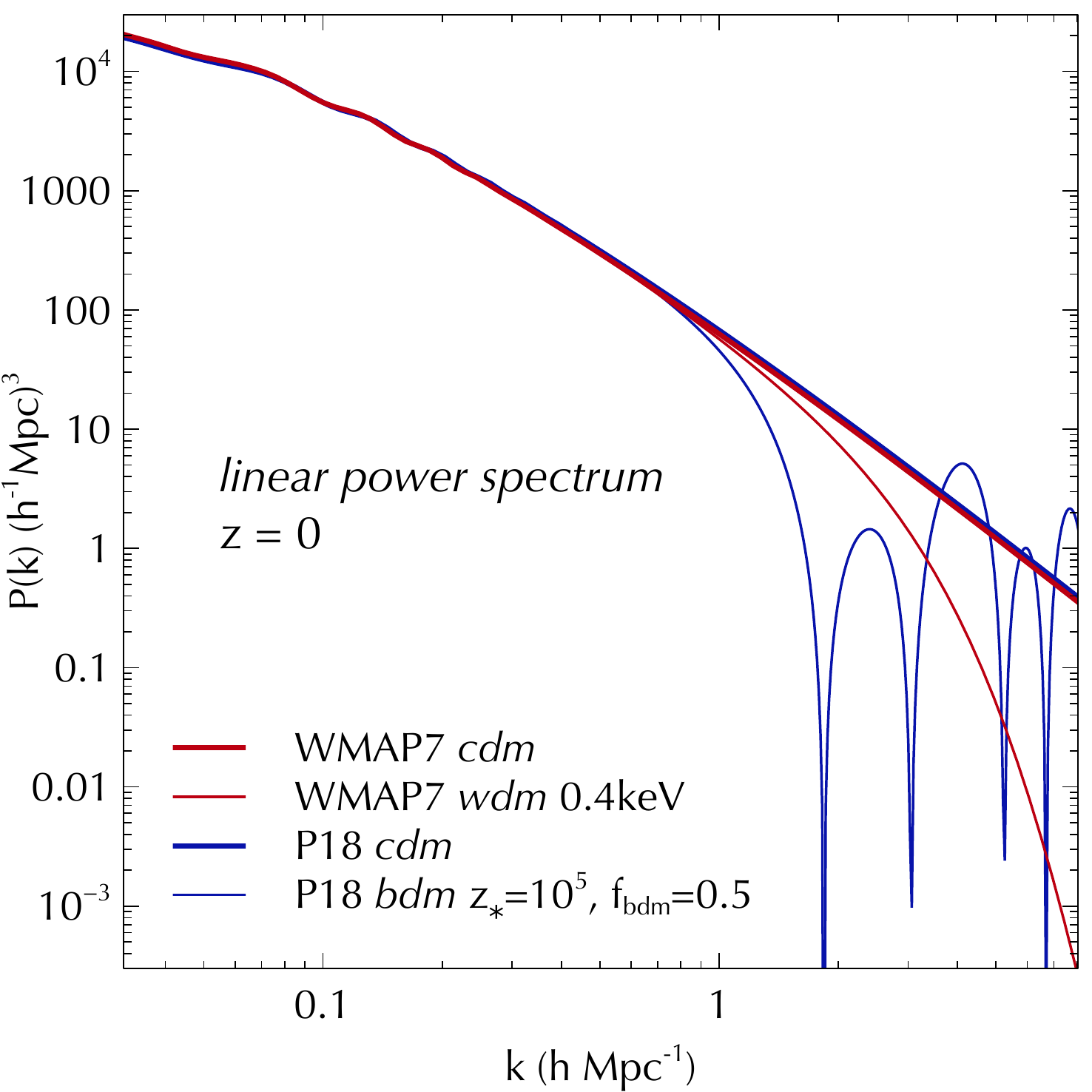}
\caption{{\bf Linear power spectra} (shown extrapolated to $z=0$) used for generating initial conditions for the WMAP7 (red) and P18 (blue) cosmologies, computed using the codes \textsc{camb} and \textsc{class}, respectively. Thick curves correspond to CDM spectra while thin curves show the WDM (red, WMAP7) and BDM (blue, P18) models used in this work. See text for a description of these models and the parameters defining them.}
\label{fig:Pklin}
\end{figure}

\begin{figure*}
\centering
\includegraphics[width=0.8\textwidth]{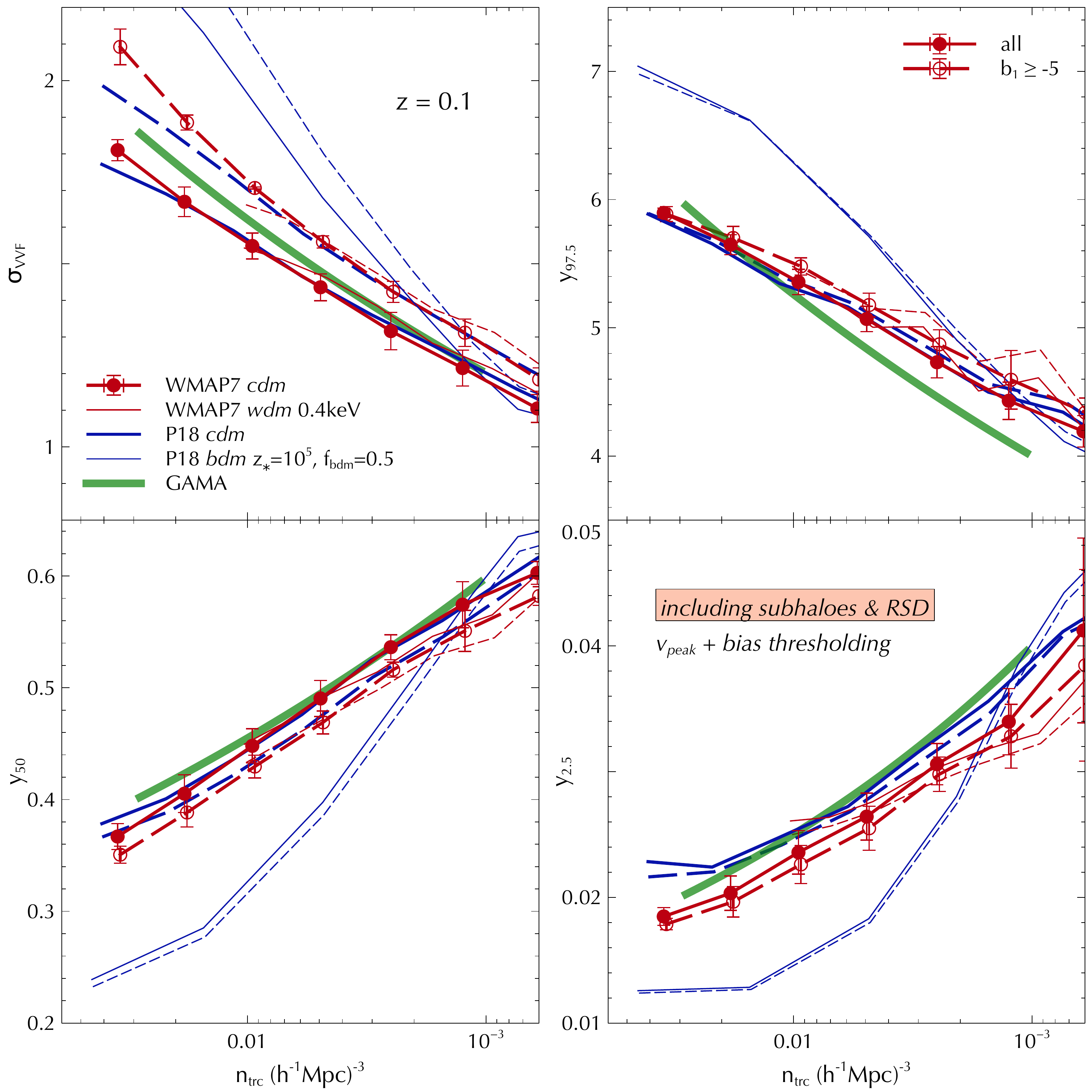}
\caption{{\bf Comparing cosmological models:} 
Similar to Figure~\ref{fig:vvf-vpeak-GAMA}, showing VVF standard deviation $\sigma_{\rm VVF}$ \emph{(top left)} and the percentiles $y_{97.5},y_{50},y_{2.5}$ (respectively, \emph{top right}, \emph{bottom left}, \emph{bottom right}) for different cosmological models: WMAP7 (red) CDM (thick lines with symbols) and WDM (thin lines), and P18 (blue) CDM (thick lines) and BDM (thin lines). See Figure~\ref{fig:Pklin} for the initial matter power spectra of these models.
The all-(sub)halo results are shown as solid lines and results for $b_1\geq-5$ as dashed lines. 
Thick solid green curves show the power-law fits to GAMA measurements from Table~\ref{tab:GAMA}.
WMAP7 CDM results were averaged over 2 realisations of the ${\rm L}150\_{\rm N}1024$ configuration, with error bars showing the standard deviation across the realisations.
We see that the BDM model for our choice of parameter values is ruled out by the GAMA measurements. See text for a discussion.}
\label{fig:vvfpercstd-vpeak-comparecosmo}
\end{figure*}

Thus, both our choices of non-standard dark matter models reflect extreme situations that will help exemplify the sensitivity of the VVF to such physics. Figure~\ref{fig:Pklin} shows the linear theory matter power spectra in these models (extrapolated to $z=0$) which were used to generate the initial conditions of our simulations. The WMAP7 (P18) CDM transfer functions were generated using \textsc{camb} (\textsc{class}). The WDM transfer function was generated using the fitting function \eqref{eq:Tk-wdm} as described above. The BDM transfer function (along with its CDM counterpart) was kindly provided by Anirban Das. For our choice of parameter values, at scales with $k\gtrsim1h/{\rm Mpc}$ this model shows peaks of alternating heights, with an overall average power that is larger than that in CDM.

The halo mass functions for the WDM and BDM cosmologies at $z=0$ are compared with their CDM counterparts in Figure~\ref{fig:hmf}. As noted earlier, at high masses all cosmologies behave similarly. At lower masses, the WDM model displays a characteristic suppression of number counts below the half-mode mass scale \citep[e.g.,][]{Schneider2012}; the origin of this suppression can be understood using peaks theory and the excursion set formalism \citep{hp14}. The BDM model, on the other hand, shows a dip near the mass scale associated with the first oscillation in its initial power spectrum, with an enhancement compared to CDM at lower masses, corresponding to the enhanced average power at high $k$ in the initial conditions. 

Collisionless $N$-body simulations with a truncation of initial power are known to be plagued by numerical artefacts (essentially, discreteness noise which `gravitates' identically to real density fluctuations) leading to a large number of spurious objects at mass scales substantially smaller than the half-mode mass \citep{Wang2007}. Careful treatments of these effects, using both traditional $N$-body techniques \citep[see, e.g.,][]{Lovell2012} as well as alternate phase space tessellation techniques \citep{shh12,hak13,Angulo2013} have led to well-calibrated mass functions for WDM cosmologies \citep{ssr13}. Figure~\ref{fig:hmf} shows that these effects are not larger than about $\sim10\%$ for our WDM model (compare the simulated mass function with the fit from the literature) over the range of mass scales we consider, owing to a combination of our virial cleaning criterion \citep[section~\ref{sec:sims}; see also][]{ac15} and the fact that we do not reach mass scales substantially below the half-mode mass. In the case of BDM, there are no similarly reliable fitting functions currently available. However, considering that this model has, on average, \emph{more} initial power than CDM at the smallest resolved scales leads us to expect that BDM-like cosmologies are likely much more robust to discreteness artefacts than WDM ones. We leave a fuller investigation of the convergence properties to future work, noting however that previous simulations with oscillatory initial power spectra have produced mass function shapes not dissimilar to the one seen in Figure~\ref{fig:hmf} \citep[][see, e.g., Figure~2 of the latter]{camd17,bose+19,sameie+19}.

Figure~\ref{fig:vvfpercstd-vpeak-comparecosmo} shows the corresponding VVF results for the all-(sub)halo samples and for samples selected by $b_1\geq-5$. 
Comparing the two CDM cosmologies with each other, we see that $\sigma_{\rm VVF}$ and $y_{50}$ are very insensitive to cosmological parameters, while the extreme percentiles $y_{97.5}$ and especially $y_{2.5}$ show larger differences, which we also saw in Figure~\ref{fig:vvf-vpeak-GAMA}. The WDM model produces nearly the same results as its CDM counterpart for all the statistics for the same choice of sample selection.
Finally, the BDM results are dramatically different from those of the other cosmologies, for both choices of sample selection, for each statistic in both amplitude and slope. 

Based on our earlier discussion, the BDM model with our choice of parameters clearly produces (sub)haloes that are far too strongly clustered compared to GAMA galaxies, when chosen by simple abundance matching. It would be very interesting to simultaneously explore the space of dark matter and astrophysical parameters to produce joint constraints. 
For now, we simply conclude that SHAM with $v_{\rm peak}$ for our choice of parameters for the BDM model dramatically fails at matching the VVF measured in the GAMA survey (downsampling and/or masking would not affect this conclusion; see Figures~\ref{fig:vvfpercstd-down-RSDsub} and~\ref{fig:vvfpercstd-mask-RSDsub}). This suggests that the VVF is very sensitive to small-scale features such as oscillations in the initial matter power spectrum.
\footnote{We do note that, at present, we cannot rule out that a traditional 2-point clustering analysis might lead to the same conclusions regarding our chosen BDM model. We will explore this issue further in future work.}

On the contrary, the WDM results for the VVF statistics in Figure~\ref{fig:vvfpercstd-vpeak-comparecosmo} are nearly identical to their CDM counterparts. Within the space of dark matter models restricted to those with strong suppression of initial power at small scales, therefore, the low-redshift VVF is evidently much more sensitive to galaxy evolution than it is to cosmology. The combination of these effects could potentially lead to interesting constraints in both of these fields. We will pursue these ideas further in future work.

\section{Discussion \& Conclusion}
\label{sec:conclude}
\noindent
We have studied the cosmological Voronoi volume function (VVF) -- the distribution of cell volumes in the Voronoi tessellation of a set of cosmological tracers. Although the VVF has appeared in the large-scale structure literature previously \citep[e.g.,][]{vdw94}, its dependence on realistic tracer properties appears to have been ignored to date. In this exploratory work, we investigated in detail the non-linear cosmological information content of the VVF in $N$-body simulations, using samples selected by a variety of tracer properties.

Intimately connected with the void probability function, the VVF of any set of clustered tracers depends on the entire infinite hierarchy of tracer correlation functions, as we showed in section~\ref{sec:vvf:theory} (see also Appendix~\ref{app:negbin}). 
Although this does not necessarily imply complete information regarding tracer properties \citep[see, e.g.,][]{cn12}, we demonstrated that
the VVF is sensitive to a variety of physical properties of the chosen sample, including halo mass, large-scale environment, the presence of substructure and redshift space distortions (section~\ref{sec:vvf:sims}), as well as selection effects such as downsampling, survey masks and halo assembly bias (Appendix~\ref{app:downmaskassembly}), all of which we explored using a suite of $N$-body simulations (section~\ref{sec:sims}). In this work, we characterised the large-scale environment of (sub)haloes using the linear bias $b_1$ of individual objects \citep[][see section~\ref{subsubsec:b1} and Appendix~\ref{app:hbyhbias}]{phs18a}.

The physical properties mentioned above are relevant not only for primordial cosmology but also for galaxy evolution. E.g., halo abundances and clustering (or environment) respond to the shape of the initial matter power spectrum and are in turn relevant for determining the nature of gas accretion eventually leading to the promotion or cessation of star formation activity in galaxies. Underdense environments tend to host relatively isolated, star forming galaxies while, at the other extreme, cluster environments tend to host `red and dead' satellite galaxies residing in subhaloes. The role of filamentary environments of the cosmic web in determining star formation activity is also of great interest, as is building a deeper understanding of the dichotomy between the evolution of central and satellite galaxies \citep[see][for a review]{sd15}. The sensitivity of the VVF to (sub)halo properties and environment therefore holds considerable promise for studying both cosmology and galaxy evolution.

We argued that the shape of the VVF is conveniently characterised by its width $\sigma_{\rm VVF}$, together with the median and a low percentile such as $y_{2.5}$ (i.e., the 2.5 percentile of $y=V/\avg{V}=n_{\rm trc}V$) as a function of tracer number density $n_{\rm trc}$. These carry complementary information on the physical attributes of the tracer sample; whereas $\sigma_{\rm VVF}$ and the median $y_{50}$ respond to large-scale bias $b_1$ in opposite ways (Figures~\ref{fig:vvfntrcstdperc-bias}, \ref{fig:vvfpercstd-bias-RSDsub} and~\ref{fig:vvf-vpeak-GAMA}), $y_{2.5}$ is particularly sensitive to the presence of substructure and redshift space distortions (Figure~\ref{fig:vvfpercstd-RSDsub-CDM}) while being very insensitive to $b_1$ (Figure~\ref{fig:vvfpercstd-bias-RSDsub}). We also showed that these observables respond smoothly to selection effects such as uniform downsampling (Figures~\ref{fig:vvfpercstd-down} and~\ref{fig:vvfpercstd-down-RSDsub}) while being essentially unaffected by the presence of survey masks (Figure~\ref{fig:vvfpercstd-mask-RSDsub}).

Our analysis of these observables has led to three main results which we summarise below.
\begin{itemize}
\item \emph{Universality in mass-selected samples:} First, within the context of flat $\Lambda$CDM cosmologies, we find that $\sigma_{\rm VVF}$ and all percentiles are approximately universal functions of $n_{\rm trc}$ for mass-thresholded samples, being very insensitive to cosmological parameters for samples chosen both with and without substructure and in real or redshift space (Figures~\ref{fig:vvfperc-ntrc}, \ref{fig:vvfstd-ntrc} and~\ref{fig:vvfpercstd-vpeak-comparecosmo}). The lower percentiles, especially $y_{2.5}$, show a strong redshift evolution which is nevertheless universal across different cosmological parameters (Figure~\ref{fig:vvfperc-ntrc}). The VVF of \emph{mass-selected samples} in flat $\Lambda$CDM is therefore much more sensitive to the physical nature of the sample, such as mean halo bias or the presence of substructure, than it is to cosmology.

\item \emph{Realistic samples and beyond-mass effects:} 
In section~\ref{sec:GAMA}, we attempted to reproduce measurements of $\sigma_{\rm VVF}$, $y_{2.5}$ and the median $y_{50}$ for luminosity-thresholded galaxy samples in the publicly available G15 field of the GAMA survey, using subhalo abundance matching (SHAM) in our CDM simulations. We showed that SHAM in redshift space with the $v_{\rm peak}$ variable at $z=0.1$ (the median redshift of GAMA) leads to reasonable descriptions of $y_{2.5}(n_{\rm trc})$ and $y_{50}(n_{\rm trc})$ using the P18 CDM cosmology, but underestimates $\sigma_{\rm VVF}(n_{\rm trc})$ (Figure~\ref{fig:vvf-vpeak-GAMA}), while the WMAP7 CDM cosmology is mildly disfavoured by $\sigma_{\rm VVF}$ as well as $y_{2.5}$. 

The mismatch with $\sigma_{\rm VVF}$ in the P18 (as well as WMAP7) simulations can be ameliorated by additionally selecting on $b_1$; e.g., we find that setting $b_1\geq-5$ simultaneously with a $v_{\rm peak}$ threshold leads to a reasonable description of $\sigma_{\rm VVF}$ while not affecting $y_{2.5}$, but mildly worsens the agreement with $y_{50}$. It is also conceivable that the $\sigma_{\rm VVF}$ mismatch arises due to differences in the properties of satellites in the observed sample and subhaloes in the simulation. In either case, the combination of VVF percentiles and standard deviation for realistically constructed (sub)halo samples is \emph{simultaneously sensitive to both cosmology and galaxy evolution.} This is particularly relevant from the point of view of beyond-mass effects in galaxy evolution, which have traditionally been challenging to quantify in observed samples \citep[see, e.g.,][]{zentner+16,zehavi+18,phs18b,azpm19,vh19,kt19} and might be better constrained by the VVF. 

\item \emph{Cosmology dependence.} Finally, we have also explored the sensitivity of the VVF to dark matter physics using simulations of two non-standard dark matter models (section~\ref{sec:cosmo}). One of these presents a small-scale truncation (WDM) and the other small-scale oscillations BDM in the initial matter power spectrum (Figure~\ref{fig:Pklin}), with model parameters deliberately set to rather extreme values. 

Our analysis showed that the truncation of initial power in the WDM model leads to essentially no imprint in VVF statistics as a function of $n_{\rm trc}$ for mass scales not much smaller than the WDM `half-mode' mass. Studying the effect at lower masses is a very interesting exercise but would require WDM simulations with much better control on numerical artefacts \citep[e.g.,][]{Angulo2013}, which is beyond the scope of this work. On the contrary, the small-scale oscillations of initial power in the BDM model lead to dramatic effects in \emph{all the VVF statistics} we studied, by substantially enhancing the clustering of tracers. Imposing further selections by environment, as above, only worsens the resulting mismatch with the observed VVF statistics of GAMA galaxies. Thus, the VVF promises to be a sensitive probe of oscillatory features in the primordial power spectrum.
\end{itemize}

Our work can be extended in a number of interesting directions. In Paper-II, we perform several halo occupation distribution (HOD) analyses on GAMA galaxies, using the percentiles and standard deviation of the VVF  as observational constraints in addition to the luminosity and 2-point correlation functions, and also exploring the role of environment beyond halo mass. Along these lines, it will also be very interesting to incorporate non-standard dark matter phenomenology to obtain joint constraints on dark matter and galaxy evolution parameters, particularly in the context of the primordial oscillatory features discussed above. 
This would require the use of fast techniques for modelling halo distributions in such models \citep[e.g.,][]{pinocchio,tze13,kh12,pinocchio-reloaded}, possibly building on the emulator framework \citep[e.g.,][]{heitmann+16,knabenhans+19}.

Our results in section~\ref{subsubsec:sub} and~\ref{subsubsec:RSD} show that low percentiles of the VVF such as $y_{2.5}$ are extremely sensitive to substructure properties and potentially also to dynamics in filamentary environments. This could be very interesting for group finder algorithms which seek to robustly separate satellite and central galaxy populations, accounting for both backsplash objects as well as observational interlopers. E.g., it would be interesting to see whether subhaloes preferentially occupy certain low percentiles of the VVF. 
In parallel, the response of the low percentiles to small-scale bulk flows (e.g., near the nodes of thick filaments) could be of interest in studies of tidally aligned substructure accretion and growth \citep[c.f., e.g.,][]{shi+20}.
These results are also potentially of interest for modelling the cosmological dependence of redshift space effects.

Finally, the strong redshift evolution of the low percentiles at large $n_{\rm trc}$ (e.g., Figures~\ref{fig:vvfperc-ntrc} and~\ref{fig:vvf-vpeak-GAMA}) deserves a more careful analysis and could potentially aid in constraining the nature of dark energy. We will return to all these issues in future work.

\section*{Acknowledgments}
We thank the Munich Institute for Astro- and Particle Physics (MIAPP) and the organisers of the programme on Dynamics of Large-Scale Structure (July 2019), where the idea for this project first emerged.
The MIAPP is funded by the Deutsche Forschungsgemeinschaft (DFG, German Research Foundation) under Germany's Excellence Strategy – EXC-2094 – 390783311.
We thank Surhud More and Sujatha Ramakrishnan for useful discussions, and our referee Mark Neyrinck for a helpful report.
The research of AP is supported by the Associateship Scheme of ICTP, Trieste and the Ramanujan Fellowship awarded by the Department of Science and Technology, Government of India. 
SA is supported by the European Research Council through the COSFORM research grant (\#670193).
This work used the open source computing packages \textsc{numpy}  \citep{vanderwalt-numpy}\footnote{\href{http://www.numpy.org}{http://www.numpy.org}}, \textsc{scipy} \citep{scipy} and the plotting software \textsc{veusz}.\footnote{\href{https://veusz.github.io/}{https://veusz.github.io/}}
We gratefully acknowledge the use of high performance computing facilities at IUCAA, Pune.

\bibliography{masterRef}

\begin{thebibliography}{}
\makeatletter
\relax
\def\mn@urlcharsother{\let\do\@makeother \do\$\do\&\do\#\do\^\do\_\do\%\do\~}
\def\mn@doi{\begingroup\mn@urlcharsother \@ifnextchar [ {\mn@doi@}
  {\mn@doi@[]}}
\def\mn@doi@[#1]#2{\def\@tempa{#1}\ifx\@tempa\@empty \href
  {http://dx.doi.org/#2} {doi:#2}\else \href {http://dx.doi.org/#2} {#1}\fi
  \endgroup}
\def\mn@eprint#1#2{\mn@eprint@#1:#2::\@nil}
\def\mn@eprint@arXiv#1{\href {http://arxiv.org/abs/#1} {{\tt arXiv:#1}}}
\def\mn@eprint@dblp#1{\href {http://dblp.uni-trier.de/rec/bibtex/#1.xml}
  {dblp:#1}}
\def\mn@eprint@#1:#2:#3:#4\@nil{\def\@tempa {#1}\def\@tempb {#2}\def\@tempc
  {#3}\ifx \@tempc \@empty \let \@tempc \@tempb \let \@tempb \@tempa \fi \ifx
  \@tempb \@empty \def\@tempb {arXiv}\fi \@ifundefined
  {mn@eprint@\@tempb}{\@tempb:\@tempc}{\expandafter \expandafter \csname
  mn@eprint@\@tempb\endcsname \expandafter{\@tempc}}}

\bibitem[\protect\citeauthoryear{{Agarwal} \& {Corasaniti}}{{Agarwal} \&
  {Corasaniti}}{2015}]{ac15}
{Agarwal} S.,  {Corasaniti} P.~S.,  2015, \mn@doi [\prd]
  {10.1103/PhysRevD.91.123509}, \href
  {https://ui.adsabs.harvard.edu/abs/2015PhRvD..91l3509A} {91, 123509}

\bibitem[\protect\citeauthoryear{{Alam}, {Zu}, {Peacock}  \&
  {Mandelbaum}}{{Alam} et~al.}{2019}]{azpm19}
{Alam} S.,  {Zu} Y.,  {Peacock} J.~A.,   {Mandelbaum} R.,  2019, \mn@doi
  [\mnras] {10.1093/mnras/sty3477}, \href
  {https://ui.adsabs.harvard.edu/\#abs/2019MNRAS.483.4501A} {483, 4501}

\bibitem[\protect\citeauthoryear{{Allgood}, {Flores}, {Primack}, {Kravtsov},
  {Wechsler}, {Faltenbacher}  \& {Bullock}}{{Allgood}
  et~al.}{2006}]{allgood+06}
{Allgood} B.,  {Flores} R.~A.,  {Primack} J.~R.,  {Kravtsov} A.~V.,  {Wechsler}
  R.~H.,  {Faltenbacher} A.,   {Bullock} J.~S.,  2006, \mn@doi [\mnras]
  {10.1111/j.1365-2966.2006.10094.x}, \href
  {http://adsabs.harvard.edu/abs/2006MNRAS.367.1781A} {367, 1781}

\bibitem[\protect\citeauthoryear{{Amendola} et~al.,}{{Amendola}
  et~al.}{2018}]{amendola+18}
{Amendola} L.,  et~al., 2018, \mn@doi [Living Reviews in Relativity]
  {10.1007/s41114-017-0010-3}, \href
  {https://ui.adsabs.harvard.edu/abs/2018LRR....21....2A} {21, 2}

\bibitem[\protect\citeauthoryear{{Angulo}, {Hahn}  \& {Abel}}{{Angulo}
  et~al.}{2013}]{Angulo2013}
{Angulo} R.~E.,  {Hahn} O.,   {Abel} T.,  2013, \mn@doi [\mnras]
  {10.1093/mnras/stt1246}, \href
  {http://adsabs.harvard.edu/abs/2013MNRAS.434.3337A} {434, 3337}

\bibitem[\protect\citeauthoryear{{Arag{\'o}n-Calvo}, {Jones}, {van de Weygaert}
   \& {van der Hulst}}{{Arag{\'o}n-Calvo} et~al.}{2007}]{aragoncalvo07}
{Arag{\'o}n-Calvo} M.~A.,  {Jones} B.~J.~T.,  {van de Weygaert} R.,   {van der
  Hulst} J.~M.,  2007, \mn@doi [\aap] {10.1051/0004-6361:20077880}, \href
  {http://adsabs.harvard.edu/abs/2007A%26A...474..315A} {474, 315}

\bibitem[\protect\citeauthoryear{{Behroozi}, {Wechsler}  \& {Wu}}{{Behroozi}
  et~al.}{2013a}]{behroozi13-rockstar}
{Behroozi} P.~S.,  {Wechsler} R.~H.,   {Wu} H.-Y.,  2013a, \mn@doi [\apj]
  {10.1088/0004-637X/762/2/109}, \href
  {http://adsabs.harvard.edu/abs/2013ApJ...762..109B} {762, 109}

\bibitem[\protect\citeauthoryear{{Behroozi}, {Wechsler}, {Wu}, {Busha},
  {Klypin}  \& {Primack}}{{Behroozi}
  et~al.}{2013b}]{behroozi13-consistenttrees}
{Behroozi} P.~S.,  {Wechsler} R.~H.,  {Wu} H.-Y.,  {Busha} M.~T.,  {Klypin}
  A.~A.,   {Primack} J.~R.,  2013b, \mn@doi [\apj]
  {10.1088/0004-637X/763/1/18}, \href
  {http://adsabs.harvard.edu/abs/2013ApJ...763...18B} {763, 18}

\bibitem[\protect\citeauthoryear{{Bernardeau} \& {van de
  Weygaert}}{{Bernardeau} \& {van de Weygaert}}{1996}]{bvdw96}
{Bernardeau} F.,  {van de Weygaert} R.,  1996, \mn@doi [\mnras]
  {10.1093/mnras/279.2.693}, \href
  {https://ui.adsabs.harvard.edu/abs/1996MNRAS.279..693B} {279, 693}

\bibitem[\protect\citeauthoryear{{Bernardeau}, {Colombi}, {Gazta{\~n}aga}  \&
  {Scoccimarro}}{{Bernardeau} et~al.}{2002}]{bcgs02}
{Bernardeau} F.,  {Colombi} S.,  {Gazta{\~n}aga} E.,   {Scoccimarro} R.,  2002,
  \mn@doi [\physrep] {10.1016/S0370-1573(02)00135-7}, \href
  {http://adsabs.harvard.edu/abs/2002PhR...367....1B} {367, 1}

\bibitem[\protect\citeauthoryear{{Bett}, {Eke}, {Frenk}, {Jenkins}, {Helly}  \&
  {Navarro}}{{Bett} et~al.}{2007}]{bett+07}
{Bett} P.,  {Eke} V.,  {Frenk} C.~S.,  {Jenkins} A.,  {Helly} J.,   {Navarro}
  J.,  2007, \mn@doi [\mnras] {10.1111/j.1365-2966.2007.11432.x}, \href
  {http://adsabs.harvard.edu/abs/2007MNRAS.376..215B} {376, 215}

\bibitem[\protect\citeauthoryear{{Blas}, {Lesgourgues}  \& {Tram}}{{Blas}
  et~al.}{2011}]{class-II}
{Blas} D.,  {Lesgourgues} J.,   {Tram} T.,  2011, \mn@doi [\jcap]
  {10.1088/1475-7516/2011/07/034}, \href
  {https://ui.adsabs.harvard.edu/abs/2011JCAP...07..034B} {2011, 034}

\bibitem[\protect\citeauthoryear{{Bode}, {Ostriker}  \& {Turok}}{{Bode}
  et~al.}{2001}]{Bode2001}
{Bode} P.,  {Ostriker} J.~P.,   {Turok} N.,  2001, \mn@doi [\apj]
  {10.1086/321541}, \href {http://adsabs.harvard.edu/abs/2001ApJ...556...93B}
  {556, 93}

\bibitem[\protect\citeauthoryear{{Bond}, {Cole}, {Efstathiou}  \&
  {Kaiser}}{{Bond} et~al.}{1991}]{bcek91}
{Bond} J.~R.,  {Cole} S.,  {Efstathiou} G.,   {Kaiser} N.,  1991, \mn@doi
  [\apj] {10.1086/170520}, \href
  {http://adsabs.harvard.edu/abs/1991ApJ...379..440B} {379, 440}

\bibitem[\protect\citeauthoryear{{Boots}}{{Boots}}{1974}]{boots74}
{Boots} B.~N.,  1974, Proc. Assoc. Am. Geogr, 6, 26

\bibitem[\protect\citeauthoryear{{Bose}, {Vogelsberger}, {Zavala}, {Pfrommer},
  {Cyr-Racine}, {Bohr}  \& {Bringmann}}{{Bose} et~al.}{2019}]{bose+19}
{Bose} S.,  {Vogelsberger} M.,  {Zavala} J.,  {Pfrommer} C.,  {Cyr-Racine}
  F.-Y.,  {Bohr} S.,   {Bringmann} T.,  2019, \mn@doi [\mnras]
  {10.1093/mnras/stz1276}, \href
  {https://ui.adsabs.harvard.edu/abs/2019MNRAS.487..522B} {487, 522}

\bibitem[\protect\citeauthoryear{{Brostow}, {Dussault}  \& {Fox}}{{Brostow}
  et~al.}{1978}]{bdf78}
{Brostow} W.,  {Dussault} J.-P.,   {Fox} B.~L.,  1978, \mn@doi [Journal of
  Computational Physics] {10.1016/0021-9991(78)90110-9}, \href
  {https://ui.adsabs.harvard.edu/abs/1978JCoPh..29...81B} {29, 81}

\bibitem[\protect\citeauthoryear{{Campbell}, {van den Bosch}, {Padmanabhan},
  {Mao}, {Zentner}, {Lange}, {Jiang}  \& {Villarreal}}{{Campbell}
  et~al.}{2018}]{campbell+18}
{Campbell} D.,  {van den Bosch} F.~C.,  {Padmanabhan} N.,  {Mao} Y.-Y.,
  {Zentner} A.~R.,  {Lange} J.~U.,  {Jiang} F.,   {Villarreal} A.,  2018,
  \mn@doi [\mnras] {10.1093/mnras/sty495}, \href
  {https://ui.adsabs.harvard.edu/abs/2018MNRAS.477..359C} {477, 359}

\bibitem[\protect\citeauthoryear{{Carretero}, {Castander}, {Gazta{\~n}aga},
  {Crocce}  \& {Fosalba}}{{Carretero} et~al.}{2015}]{carretero+15}
{Carretero} J.,  {Castander} F.~J.,  {Gazta{\~n}aga} E.,  {Crocce} M.,
  {Fosalba} P.,  2015, \mn@doi [\mnras] {10.1093/mnras/stu2402}, \href
  {http://adsabs.harvard.edu/abs/2015MNRAS.447..646C} {447, 646}

\bibitem[\protect\citeauthoryear{{Carron} \& {Neyrinck}}{{Carron} \&
  {Neyrinck}}{2012}]{cn12}
{Carron} J.,  {Neyrinck} M.~C.,  2012, \mn@doi [\apj]
  {10.1088/0004-637X/750/1/28}, \href
  {https://ui.adsabs.harvard.edu/abs/2012ApJ...750...28C} {750, 28}

\bibitem[\protect\citeauthoryear{{Coles}}{{Coles}}{1990}]{coles90}
{Coles} P.,  1990, \mn@doi [\nat] {10.1038/346446a0}, \href
  {https://ui.adsabs.harvard.edu/abs/1990Natur.346..446C} {346, 446}

\bibitem[\protect\citeauthoryear{{Corasaniti}, {Agarwal}, {Marsh}  \&
  {Das}}{{Corasaniti} et~al.}{2017}]{camd17}
{Corasaniti} P.~S.,  {Agarwal} S.,  {Marsh} D.~J.~E.,   {Das} S.,  2017,
  \mn@doi [\prd] {10.1103/PhysRevD.95.083512}, \href
  {https://ui.adsabs.harvard.edu/abs/2017PhRvD..95h3512C} {95, 083512}

\bibitem[\protect\citeauthoryear{{Croton} et~al.,}{{Croton}
  et~al.}{2004}]{croton+04}
{Croton} D.~J.,  et~al., 2004, \mn@doi [\mnras]
  {10.1111/j.1365-2966.2004.07968.x}, \href
  {https://ui.adsabs.harvard.edu/abs/2004MNRAS.352..828C} {352, 828}

\bibitem[\protect\citeauthoryear{{Cyr-Racine}, {Sigurdson}, {Zavala},
  {Bringmann}, {Vogelsberger}  \& {Pfrommer}}{{Cyr-Racine}
  et~al.}{2016}]{cyr-racine+16}
{Cyr-Racine} F.-Y.,  {Sigurdson} K.,  {Zavala} J.,  {Bringmann} T.,
  {Vogelsberger} M.,   {Pfrommer} C.,  2016, \mn@doi [\prd]
  {10.1103/PhysRevD.93.123527}, \href
  {https://ui.adsabs.harvard.edu/abs/2016PhRvD..93l3527C} {93, 123527}

\bibitem[\protect\citeauthoryear{{Das}, {Dasgupta}  \& {Khatri}}{{Das}
  et~al.}{2019}]{ddk19}
{Das} A.,  {Dasgupta} B.,   {Khatri} R.,  2019, \mn@doi [\jcap]
  {10.1088/1475-7516/2019/04/018}, \href
  {https://ui.adsabs.harvard.edu/abs/2019JCAP...04..018D} {2019, 018}

\bibitem[\protect\citeauthoryear{Dirichlet}{Dirichlet}{1850}]{dirichlet1850}
Dirichlet G.~L.,  1850, J. Reine Angew. Math., 40, 209

\bibitem[\protect\citeauthoryear{{Driver} et~al.,}{{Driver}
  et~al.}{2009}]{driver+09}
{Driver} S.~P.,  et~al., 2009, \mn@doi [Astronomy and Geophysics]
  {10.1111/j.1468-4004.2009.50512.x}, \href
  {https://ui.adsabs.harvard.edu/abs/2009A&G....50e..12D} {50, 5.12}

\bibitem[\protect\citeauthoryear{{Elizalde} \& {Gaztanaga}}{{Elizalde} \&
  {Gaztanaga}}{1992}]{eg92}
{Elizalde} E.,  {Gaztanaga} E.,  1992, \mn@doi [\mnras]
  {10.1093/mnras/254.2.247}, \href
  {https://ui.adsabs.harvard.edu/abs/1992MNRAS.254..247E} {254, 247}

\bibitem[\protect\citeauthoryear{{Fall}, {Geller}, {Jones}  \& {White}}{{Fall}
  et~al.}{1976}]{fgjw76}
{Fall} S.~M.,  {Geller} M.~J.,  {Jones} B. J.~T.,   {White} S. D.~M.,  1976,
  \mn@doi [\apjl] {10.1086/182104}, \href
  {https://ui.adsabs.harvard.edu/abs/1976ApJ...205L.121F} {205, L121}

\bibitem[\protect\citeauthoryear{{Faltenbacher} \& {White}}{{Faltenbacher} \&
  {White}}{2010}]{fw10}
{Faltenbacher} A.,  {White} S.~D.~M.,  2010, \mn@doi [\apj]
  {10.1088/0004-637X/708/1/469}, \href
  {http://adsabs.harvard.edu/abs/2010ApJ...708..469F} {708, 469}

\bibitem[\protect\citeauthoryear{{Ferenc} \& {N{\'e}da}}{{Ferenc} \&
  {N{\'e}da}}{2007}]{fn07}
{Ferenc} J.-S.,  {N{\'e}da} Z.,  2007, \mn@doi [Physica A Statistical Mechanics
  and its Applications] {10.1016/j.physa.2007.07.063}, \href
  {https://ui.adsabs.harvard.edu/abs/2007PhyA..385..518F} {385, 518}

\bibitem[\protect\citeauthoryear{{Fry}}{{Fry}}{1986}]{fry86}
{Fry} J.~N.,  1986, \mn@doi [\apj] {10.1086/164348}, \href
  {https://ui.adsabs.harvard.edu/abs/1986ApJ...306..358F} {306, 358}

\bibitem[\protect\citeauthoryear{{Fry} \& {Colombi}}{{Fry} \&
  {Colombi}}{2013}]{fc13}
{Fry} J.~N.,  {Colombi} S.,  2013, \mn@doi [\mnras] {10.1093/mnras/stt745},
  \href {https://ui.adsabs.harvard.edu/abs/2013MNRAS.433..581F} {433, 581}

\bibitem[\protect\citeauthoryear{{Gao}, {Springel}  \& {White}}{{Gao}
  et~al.}{2005}]{gsw05}
{Gao} L.,  {Springel} V.,   {White} S.~D.~M.,  2005, \mn@doi [\mnras]
  {10.1111/j.1745-3933.2005.00084.x}, \href
  {http://adsabs.harvard.edu/abs/2005MNRAS.363L..66G} {363, L66}

\bibitem[\protect\citeauthoryear{{Garzilli}, {Ruchayskiy}, {Magalich}  \&
  {Boyarsky}}{{Garzilli} et~al.}{2019}]{grmb20}
{Garzilli} A.,  {Ruchayskiy} O.,  {Magalich} A.,   {Boyarsky} A.,  2019, arXiv
  e-prints, \href {https://ui.adsabs.harvard.edu/abs/2019arXiv191209397G} {p.
  arXiv:1912.09397}

\bibitem[\protect\citeauthoryear{{Gerke}, {Wechsler}, {Behroozi}, {Cooper},
  {Yan}  \& {Coil}}{{Gerke} et~al.}{2013}]{gerke+13}
{Gerke} B.~F.,  {Wechsler} R.~H.,  {Behroozi} P.~S.,  {Cooper} M.~C.,  {Yan}
  R.,   {Coil} A.~L.,  2013, \mn@doi [\apjs] {10.1088/0067-0049/208/1/1}, \href
  {https://ui.adsabs.harvard.edu/abs/2013ApJS..208....1G} {208, 1}

\bibitem[\protect\citeauthoryear{Gilbert}{Gilbert}{1962}]{gilbert62}
Gilbert E.~N.,  1962, \mn@doi [Ann. Math. Statist.] {10.1214/aoms/1177704464},
  33, 958

\bibitem[\protect\citeauthoryear{{Gill}, {Knebe}  \& {Gibson}}{{Gill}
  et~al.}{2005}]{gkg05}
{Gill} S.~P.~D.,  {Knebe} A.,   {Gibson} B.~K.,  2005, \mn@doi [\mnras]
  {10.1111/j.1365-2966.2004.08562.x}, \href
  {http://adsabs.harvard.edu/abs/2005MNRAS.356.1327G} {356, 1327}

\bibitem[\protect\citeauthoryear{{Hahn} \& {Abel}}{{Hahn} \&
  {Abel}}{2011}]{hahn11-music}
{Hahn} O.,  {Abel} T.,  2011, \mn@doi [\mnras]
  {10.1111/j.1365-2966.2011.18820.x}, \href
  {http://adsabs.harvard.edu/abs/2011MNRAS.415.2101H} {415, 2101}

\bibitem[\protect\citeauthoryear{{Hahn} \& {Paranjape}}{{Hahn} \&
  {Paranjape}}{2014}]{hp14}
{Hahn} O.,  {Paranjape} A.,  2014, \mn@doi [\mnras] {10.1093/mnras/stt2256},
  \href {http://adsabs.harvard.edu/abs/2014MNRAS.438..878H} {438, 878}

\bibitem[\protect\citeauthoryear{{Hahn}, {Abel}  \& {Kaehler}}{{Hahn}
  et~al.}{2013}]{hak13}
{Hahn} O.,  {Abel} T.,   {Kaehler} R.,  2013, \mn@doi [\mnras]
  {10.1093/mnras/stt1061}, \href
  {http://adsabs.harvard.edu/abs/2013MNRAS.tmp.1799H} {}

\bibitem[\protect\citeauthoryear{{Hahn}, {Angulo}  \& {Abel}}{{Hahn}
  et~al.}{2015}]{haa15}
{Hahn} O.,  {Angulo} R.~E.,   {Abel} T.,  2015, \mn@doi [\mnras]
  {10.1093/mnras/stv2179}, \href
  {https://ui.adsabs.harvard.edu/abs/2015MNRAS.454.3920H} {454, 3920}

\bibitem[\protect\citeauthoryear{{Heitmann} et~al.,}{{Heitmann}
  et~al.}{2016}]{heitmann+16}
{Heitmann} K.,  et~al., 2016, \mn@doi [\apj] {10.3847/0004-637X/820/2/108},
  \href {https://ui.adsabs.harvard.edu/abs/2016ApJ...820..108H} {820, 108}

\bibitem[\protect\citeauthoryear{{Icke} \& {van de Weygaert}}{{Icke} \& {van de
  Weygaert}}{1987}]{ivdw87}
{Icke} V.,  {van de Weygaert} R.,  1987, \aap, \href
  {https://ui.adsabs.harvard.edu/abs/1987A&A...184...16I} {184, 16}

\bibitem[\protect\citeauthoryear{{Ir{\v{s}}i{\v{c}}}
  et~al.,}{{Ir{\v{s}}i{\v{c}}} et~al.}{2017}]{irsic+17}
{Ir{\v{s}}i{\v{c}}} V.,  et~al., 2017, \mn@doi [\prd]
  {10.1103/PhysRevD.96.023522}, \href
  {https://ui.adsabs.harvard.edu/abs/2017PhRvD..96b3522I} {96, 023522}

\bibitem[\protect\citeauthoryear{Jones, Oliphant, Peterson  et~al.}{Jones
  et~al.}{01  }]{scipy}
Jones E.,  Oliphant T.,  Peterson P.,   et~al., 2001--, {SciPy}: Open source
  scientific tools for {Python}, \url {http://www.scipy.org/}

\bibitem[\protect\citeauthoryear{{Kaiser}}{{Kaiser}}{1987}]{kaiser87}
{Kaiser} N.,  1987, \mn@doi [\mnras] {10.1093/mnras/227.1.1}, \href
  {https://ui.adsabs.harvard.edu/abs/1987MNRAS.227....1K} {227, 1}

\bibitem[\protect\citeauthoryear{{Kiang}}{{Kiang}}{1966}]{kiang66}
{Kiang} T.,  1966, \zap, \href
  {https://ui.adsabs.harvard.edu/abs/1966ZA.....64..433K} {64, 433}

\bibitem[\protect\citeauthoryear{{Kitaura} \& {He{\ss}}}{{Kitaura} \&
  {He{\ss}}}{2013}]{kh12}
{Kitaura} F.-S.,  {He{\ss}} S.,  2013, \mn@doi [\mnras]
  {10.1093/mnrasl/slt101}, \href
  {http://adsabs.harvard.edu/abs/2013MNRAS.435L..78K} {435, L78}

\bibitem[\protect\citeauthoryear{{Knabenhans} et~al.,}{{Knabenhans}
  et~al.}{2019}]{knabenhans+19}
{Knabenhans} M.,  et~al., 2019, \mn@doi [\mnras] {10.1093/mnras/stz197}, \href
  {https://ui.adsabs.harvard.edu/abs/2019MNRAS.484.5509K} {484, 5509}

\bibitem[\protect\citeauthoryear{{Komatsu} et~al.,}{{Komatsu}
  et~al.}{2011}]{Komatsu2010}
{Komatsu} E.,  et~al., 2011, \mn@doi [\apjs] {10.1088/0067-0049/192/2/18},
  \href {http://adsabs.harvard.edu/abs/2011ApJS..192...18K} {192, 18}

\bibitem[\protect\citeauthoryear{{Kumar}, {Kurtz}, {Banavar}  \&
  {Sharma}}{{Kumar} et~al.}{1992}]{kumar+92}
{Kumar} S.,  {Kurtz} S.~K.,  {Banavar} J.~R.,   {Sharma} M.~G.,  1992, \mn@doi
  [Journal of Statistical Physics] {10.1007/BF01049719}, \href
  {https://ui.adsabs.harvard.edu/abs/1992JSP....67..523K} {67, 523}

\bibitem[\protect\citeauthoryear{{Lesgourgues}}{{Lesgourgues}}{2011}]{class-I}
{Lesgourgues} J.,  2011, arXiv e-prints, \href
  {https://ui.adsabs.harvard.edu/abs/2011arXiv1104.2932L} {p. arXiv:1104.2932}

\bibitem[\protect\citeauthoryear{{Lewis}, {Challinor}  \& {Lasenby}}{{Lewis}
  et~al.}{2000}]{camb}
{Lewis} A.,  {Challinor} A.,   {Lasenby} A.,  2000, \mn@doi [\apj]
  {10.1086/309179}, \href {http://adsabs.harvard.edu/abs/2000ApJ...538..473L}
  {538, 473}

\bibitem[\protect\citeauthoryear{{Lovell} et~al.,}{{Lovell}
  et~al.}{2012}]{Lovell2012}
{Lovell} M.~R.,  et~al., 2012, \mn@doi [\mnras]
  {10.1111/j.1365-2966.2011.20200.x}, \href
  {http://adsabs.harvard.edu/abs/2012MNRAS.420.2318L} {420, 2318}

\bibitem[\protect\citeauthoryear{{Mart{\'\i}nez} \& {Saar}}{{Mart{\'\i}nez} \&
  {Saar}}{2002}]{ms02}
{Mart{\'\i}nez} V.~J.,  {Saar} E.,  2002, {Statistics of the Galaxy
  Distribution}.
Chapman \& Hall/CRC, Boca Raton

\bibitem[\protect\citeauthoryear{{Maurogordato} \&
  {Lachieze-Rey}}{{Maurogordato} \& {Lachieze-Rey}}{1987}]{ml-r87}
{Maurogordato} S.,  {Lachieze-Rey} M.,  1987, \mn@doi [\apj] {10.1086/165520},
  \href {https://ui.adsabs.harvard.edu/abs/1987ApJ...320...13M} {320, 13}

\bibitem[\protect\citeauthoryear{{Meijering}}{{Meijering}}{1953}]{meijering53}
{Meijering} J.~L.,  1953, Philips Res. Rep., 8, 270

\bibitem[\protect\citeauthoryear{Miles}{Miles}{1970}]{miles70}
Miles R.,  1970, \mn@doi [Mathematical Biosciences]
  {https://doi.org/10.1016/0025-5564(70)90061-1}, 6, 85

\bibitem[\protect\citeauthoryear{{Monaco}, {Theuns}  \& {Taffoni}}{{Monaco}
  et~al.}{2002}]{pinocchio}
{Monaco} P.,  {Theuns} T.,   {Taffoni} G.,  2002, \mnras, 331, 587

\bibitem[\protect\citeauthoryear{{Monaco}, {Sefusatti}, {Borgani}, {Crocce},
  {Fosalba}, {Sheth}  \& {Theuns}}{{Monaco} et~al.}{2013}]{pinocchio-reloaded}
{Monaco} P.,  {Sefusatti} E.,  {Borgani} S.,  {Crocce} M.,  {Fosalba} P.,
  {Sheth} R.~K.,   {Theuns} T.,  2013, \mnras, accepted

\bibitem[\protect\citeauthoryear{Møller}{Møller}{1989}]{moller89}
Møller J.,  1989, \mn@doi [Advances in Applied Probability] {10.2307/1427197},
  21, 37–73

\bibitem[\protect\citeauthoryear{{Neyrinck}}{{Neyrinck}}{2008}]{Neyrinck2008}
{Neyrinck} M.~C.,  2008, \mn@doi [\mnras] {10.1111/j.1365-2966.2008.13180.x},
  \href {http://adsabs.harvard.edu/abs/2008MNRAS.386.2101N} {386, 2101}

\bibitem[\protect\citeauthoryear{{Neyrinck}}{{Neyrinck}}{2013}]{neyrinck13}
{Neyrinck} M.~C.,  2013, \mn@doi [\mnras] {10.1093/mnras/sts027}, \href
  {https://ui.adsabs.harvard.edu/abs/2013MNRAS.428..141N} {428, 141}

\bibitem[\protect\citeauthoryear{{Okabe}, {Boots}  \& {Sugihara}}{{Okabe}
  et~al.}{1992}]{obs92}
{Okabe} A.,  {Boots} B.,   {Sugihara} K.,  1992, {Spatial tessellations.
  Concepts and Applications of Voronoi diagrams}.
Chichester: John Wiley

\bibitem[\protect\citeauthoryear{{Palanque-Delabrouille}, {Y{\`e}che},
  {Sch{\"o}neberg}, {Lesgourgues}, {Walther}, {Chabanier}  \&
  {Armengaud}}{{Palanque-Delabrouille} et~al.}{2019}]{pd+20}
{Palanque-Delabrouille} N.,  {Y{\`e}che} C.,  {Sch{\"o}neberg} N.,
  {Lesgourgues} J.,  {Walther} M.,  {Chabanier} S.,   {Armengaud} E.,  2019,
  arXiv e-prints, \href {https://ui.adsabs.harvard.edu/abs/2019arXiv191109073P}
  {p. arXiv:1911.09073}

\bibitem[\protect\citeauthoryear{{Paranjape}, {Hahn}  \& {Sheth}}{{Paranjape}
  et~al.}{2018a}]{phs18a}
{Paranjape} A.,  {Hahn} O.,   {Sheth} R.~K.,  2018a, \mn@doi [\mnras]
  {10.1093/mnras/sty496}, \href
  {http://adsabs.harvard.edu/abs/2018MNRAS.476.3631P} {476, 3631}

\bibitem[\protect\citeauthoryear{{Paranjape}, {Hahn}  \& {Sheth}}{{Paranjape}
  et~al.}{2018b}]{phs18b}
{Paranjape} A.,  {Hahn} O.,   {Sheth} R.~K.,  2018b, \mn@doi [\mnras]
  {10.1093/mnras/sty633}, \href
  {http://adsabs.harvard.edu/abs/2018MNRAS.476.5442P} {476, 5442}

\bibitem[\protect\citeauthoryear{{Peebles}}{{Peebles}}{1969}]{peebles69}
{Peebles} P.~J.~E.,  1969, \mn@doi [\apj] {10.1086/149876}, \href
  {http://adsabs.harvard.edu/abs/1969ApJ...155..393P} {155, 393}

\bibitem[\protect\citeauthoryear{{Peebles}}{{Peebles}}{1980}]{peebles-1980}
{Peebles} P.~J.~E.,  1980, {The large-scale structure of the universe}.
Princeton University Press, Princeton, N.J.

\bibitem[\protect\citeauthoryear{{Planck Collaboration} et~al.,}{{Planck
  Collaboration} et~al.}{2014}]{Planck13-XVI-cosmoparam}
{Planck Collaboration} et~al., 2014, \mn@doi [\aap]
  {10.1051/0004-6361/201321591}, \href
  {http://adsabs.harvard.edu/abs/2014A%26A...571A..16P} {571, A16}

\bibitem[\protect\citeauthoryear{{Planck Collaboration} et~al.,}{{Planck
  Collaboration} et~al.}{2018}]{Planck18-VI-cosmoparam}
{Planck Collaboration} et~al., 2018, arXiv e-prints, \href
  {https://ui.adsabs.harvard.edu/abs/2018arXiv180706209P} {p. arXiv:1807.06209}

\bibitem[\protect\citeauthoryear{{Platen}, {van de Weygaert}  \&
  {Jones}}{{Platen} et~al.}{2007}]{pvdwj07}
{Platen} E.,  {van de Weygaert} R.,   {Jones} B.~J.~T.,  2007, \mn@doi [\mnras]
  {10.1111/j.1365-2966.2007.12125.x}, \href
  {http://adsabs.harvard.edu/abs/2007MNRAS.380..551P} {380, 551}

\bibitem[\protect\citeauthoryear{{Press} \& {Schechter}}{{Press} \&
  {Schechter}}{1974}]{ps74}
{Press} W.~H.,  {Schechter} P.,  1974, \apj, \href
  {http://dx.doi.org/10.1086/152650} {187, 425}

\bibitem[\protect\citeauthoryear{{Ramakrishnan}, {Paranjape}, {Hahn}  \&
  {Sheth}}{{Ramakrishnan} et~al.}{2019}]{rphs19}
{Ramakrishnan} S.,  {Paranjape} A.,  {Hahn} O.,   {Sheth} R.~K.,  2019, \mn@doi
  [\mnras] {10.1093/mnras/stz2344}, \href
  {https://ui.adsabs.harvard.edu/abs/2019MNRAS.489.2977R} {489, 2977}

\bibitem[\protect\citeauthoryear{{Reddick}, {Wechsler}, {Tinker}  \&
  {Behroozi}}{{Reddick} et~al.}{2013}]{rwtb13}
{Reddick} R.~M.,  {Wechsler} R.~H.,  {Tinker} J.~L.,   {Behroozi} P.~S.,  2013,
  \mn@doi [\apj] {10.1088/0004-637X/771/1/30}, \href
  {https://ui.adsabs.harvard.edu/abs/2013ApJ...771...30R} {771, 30}

\bibitem[\protect\citeauthoryear{{Sameie}, {Benson}, {Sales}, {Yu}, {Moustakas}
   \& {Creasey}}{{Sameie} et~al.}{2019}]{sameie+19}
{Sameie} O.,  {Benson} A.~J.,  {Sales} L.~V.,  {Yu} H.-b.,  {Moustakas} L.~A.,
   {Creasey} P.,  2019, \mn@doi [\apj] {10.3847/1538-4357/ab0824}, \href
  {https://ui.adsabs.harvard.edu/abs/2019ApJ...874..101S} {874, 101}

\bibitem[\protect\citeauthoryear{{Schaap} \& {van de Weygaert}}{{Schaap} \&
  {van de Weygaert}}{2000}]{SchaapVandeWeygaert2000}
{Schaap} W.~E.,  {van de Weygaert} R.,  2000, \aap, \href
  {http://adsabs.harvard.edu/abs/2000A%26A...363L..29S} {363, L29}

\bibitem[\protect\citeauthoryear{{Schneider}, {Smith}, {Macci{\`o}}  \&
  {Moore}}{{Schneider} et~al.}{2012}]{Schneider2012}
{Schneider} A.,  {Smith} R.~E.,  {Macci{\`o}} A.~V.,   {Moore} B.,  2012,
  \mn@doi [\mnras] {10.1111/j.1365-2966.2012.21252.x}, \href
  {http://adsabs.harvard.edu/abs/2012MNRAS.424..684S} {424, 684}

\bibitem[\protect\citeauthoryear{{Schneider}, {Smith}  \& {Reed}}{{Schneider}
  et~al.}{2013}]{ssr13}
{Schneider} A.,  {Smith} R.~E.,   {Reed} D.,  2013, \mn@doi [\mnras]
  {10.1093/mnras/stt829}, \href
  {https://ui.adsabs.harvard.edu/abs/2013MNRAS.433.1573S} {433, 1573}

\bibitem[\protect\citeauthoryear{{Scoccimarro}}{{Scoccimarro}}{1998}]{scoccimarro98}
{Scoccimarro} R.,  1998, \mn@doi [\mnras] {10.1046/j.1365-8711.1998.01845.x},
  \href {http://adsabs.harvard.edu/abs/1998MNRAS.299.1097S} {299, 1097}

\bibitem[\protect\citeauthoryear{{Shandarin}, {Habib}  \&
  {Heitmann}}{{Shandarin} et~al.}{2012}]{shh12}
{Shandarin} S.,  {Habib} S.,   {Heitmann} K.,  2012, \mn@doi [\prd]
  {10.1103/PhysRevD.85.083005}, \href
  {https://ui.adsabs.harvard.edu/abs/2012PhRvD..85h3005S} {85, 083005}

\bibitem[\protect\citeauthoryear{{Sheth}}{{Sheth}}{1996}]{sheth96}
{Sheth} R.~K.,  1996, \mn@doi [\mnras] {10.1093/mnras/278.1.101}, \href
  {https://ui.adsabs.harvard.edu/abs/1996MNRAS.278..101S} {278, 101}

\bibitem[\protect\citeauthoryear{{Sheth} \& {Tormen}}{{Sheth} \&
  {Tormen}}{1999}]{st99}
{Sheth} R.~K.,  {Tormen} G.,  1999, \mn@doi [\mnras]
  {10.1046/j.1365-8711.1999.02692.x}, \href
  {http://adsabs.harvard.edu/abs/1999MNRAS.308..119S} {308, 119}

\bibitem[\protect\citeauthoryear{{Sheth} \& {Tormen}}{{Sheth} \&
  {Tormen}}{2004}]{st04}
{Sheth} R.~K.,  {Tormen} G.,  2004, \mn@doi [\mnras]
  {10.1111/j.1365-2966.2004.07733.x}, \href
  {http://adsabs.harvard.edu/abs/2004MNRAS.350.1385S} {350, 1385}

\bibitem[\protect\citeauthoryear{{Shi} et~al.,}{{Shi} et~al.}{2020}]{shi+20}
{Shi} J.,  et~al., 2020, arXiv e-prints, \href
  {https://ui.adsabs.harvard.edu/abs/2020arXiv200104090S} {p. arXiv:2001.04090}

\bibitem[\protect\citeauthoryear{{Smith}}{{Smith}}{2009}]{smith09}
{Smith} R.~E.,  2009, \mn@doi [\mnras] {10.1111/j.1365-2966.2009.15490.x},
  \href {https://ui.adsabs.harvard.edu/abs/2009MNRAS.400..851S} {400, 851}

\bibitem[\protect\citeauthoryear{{Somerville} \& {Dav{\'e}}}{{Somerville} \&
  {Dav{\'e}}}{2015}]{sd15}
{Somerville} R.~S.,  {Dav{\'e}} R.,  2015, \mn@doi [\araa]
  {10.1146/annurev-astro-082812-140951}, \href
  {http://adsabs.harvard.edu/abs/2015ARA%26A..53...51S} {53, 51}

\bibitem[\protect\citeauthoryear{{Springel}}{{Springel}}{2005}]{springel:2005}
{Springel} V.,  2005, \mn@doi [\mnras] {10.1111/j.1365-2966.2005.09655.x},
  \href {http://adsabs.harvard.edu/abs/2005MNRAS.364.1105S} {364, 1105}

\bibitem[\protect\citeauthoryear{{Springel}}{{Springel}}{2010}]{springel10}
{Springel} V.,  2010, \mn@doi [\mnras] {10.1111/j.1365-2966.2009.15715.x},
  \href {https://ui.adsabs.harvard.edu/abs/2010MNRAS.401..791S} {401, 791}

\bibitem[\protect\citeauthoryear{{Tanemura}}{{Tanemura}}{2003}]{tanemura03}
{Tanemura} M.,  2003, Forma, 18, 221

\bibitem[\protect\citeauthoryear{{Tassev}, {Zaldarriaga}  \&
  {Eisenstein}}{{Tassev} et~al.}{2013}]{tze13}
{Tassev} S.,  {Zaldarriaga} M.,   {Eisenstein} D.~J.,  2013, JCAP, 6, 36

\bibitem[\protect\citeauthoryear{{Thiessen}}{{Thiessen}}{1911}]{thiessen1911}
{Thiessen} A.~H.,  1911, \mn@doi [Monthly Weather Review]
  {10.1175/1520-0493(1911)39<1082b:PAFLA>2.0.CO;2}, \href
  {https://ui.adsabs.harvard.edu/abs/1911MWRv...39R1082T} {39, 1082}

\bibitem[\protect\citeauthoryear{{Tinker}, {Kravtsov}, {Klypin}, {Abazajian},
  {Warren}, {Yepes}, {Gottl{\"o}ber}  \& {Holz}}{{Tinker}
  et~al.}{2008}]{Tinker08}
{Tinker} J.,  {Kravtsov} A.~V.,  {Klypin} A.,  {Abazajian} K.,  {Warren} M.,
  {Yepes} G.,  {Gottl{\"o}ber} S.,   {Holz} D.~E.,  2008, \mn@doi [\apj]
  {10.1086/591439}, \href {http://adsabs.harvard.edu/abs/2008ApJ...688..709T}
  {688, 709}

\bibitem[\protect\citeauthoryear{{Tojeiro} et~al.,}{{Tojeiro}
  et~al.}{2017}]{tojeiro+17}
{Tojeiro} R.,  et~al., 2017, \mn@doi [\mnras] {10.1093/mnras/stx1466}, \href
  {http://cdsads.u-strasbg.fr/abs/2017MNRAS.470.3720T} {470, 3720}

\bibitem[\protect\citeauthoryear{{Tr{\"o}ster} et~al.,}{{Tr{\"o}ster}
  et~al.}{2019}]{troster+19}
{Tr{\"o}ster} T.,  et~al., 2019, arXiv e-prints, \href
  {https://ui.adsabs.harvard.edu/abs/2019arXiv190911006T} {p. arXiv:1909.11006}

\bibitem[\protect\citeauthoryear{{Vakili} \& {Hahn}}{{Vakili} \&
  {Hahn}}{2019}]{vh19}
{Vakili} M.,  {Hahn} C.,  2019, \mn@doi [\apj] {10.3847/1538-4357/aaf1a1},
  \href {https://ui.adsabs.harvard.edu/abs/2019ApJ...872..115V} {872, 115}

\bibitem[\protect\citeauthoryear{{Van Der Walt}, {Colbert}  \&
  {Varoquaux}}{{Van Der Walt} et~al.}{2011}]{vanderwalt-numpy}
{Van Der Walt} S.,  {Colbert} S.~C.,   {Varoquaux} G.,  2011, preprint, \href
  {http://adsabs.harvard.edu/abs/2011arXiv1102.1523V} {} (\mn@eprint {arXiv}
  {1102.1523})

\bibitem[\protect\citeauthoryear{{Viel}, {Lesgourgues}, {Haehnelt}, {Matarrese}
   \& {Riotto}}{{Viel} et~al.}{2005}]{Viel2005}
{Viel} M.,  {Lesgourgues} J.,  {Haehnelt} M.~G.,  {Matarrese} S.,   {Riotto}
  A.,  2005, \mn@doi [\prd] {10.1103/PhysRevD.71.063534}, \href
  {http://adsabs.harvard.edu/abs/2005PhRvD..71f3534V} {71, 063534}

\bibitem[\protect\citeauthoryear{{Viel}, {Becker}, {Bolton}  \&
  {Haehnelt}}{{Viel} et~al.}{2013}]{Viel2013}
{Viel} M.,  {Becker} G.~D.,  {Bolton} J.~S.,   {Haehnelt} M.~G.,  2013, \mn@doi
  [\prd] {10.1103/PhysRevD.88.043502}, \href
  {http://adsabs.harvard.edu/abs/2013PhRvD..88d3502V} {88, 043502}

\bibitem[\protect\citeauthoryear{{Vogeley}, {Geller}, {Park}  \&
  {Huchra}}{{Vogeley} et~al.}{1994}]{vgph94}
{Vogeley} M.~S.,  {Geller} M.~J.,  {Park} C.,   {Huchra} J.~P.,  1994, \mn@doi
  [\aj] {10.1086/117110}, \href
  {https://ui.adsabs.harvard.edu/abs/1994AJ....108..745V} {108, 745}

\bibitem[\protect\citeauthoryear{{Vogelsberger}, {Sijacki}, {Kere{\v{s}}},
  {Springel}  \& {Hernquist}}{{Vogelsberger} et~al.}{2012}]{vogelsberger+12}
{Vogelsberger} M.,  {Sijacki} D.,  {Kere{\v{s}}} D.,  {Springel} V.,
  {Hernquist} L.,  2012, \mn@doi [\mnras] {10.1111/j.1365-2966.2012.21590.x},
  \href {https://ui.adsabs.harvard.edu/abs/2012MNRAS.425.3024V} {425, 3024}

\bibitem[\protect\citeauthoryear{{Vogelsberger}, {Zavala}, {Cyr-Racine},
  {Pfrommer}, {Bringmann}  \& {Sigurdson}}{{Vogelsberger}
  et~al.}{2016}]{vogelsberger+16}
{Vogelsberger} M.,  {Zavala} J.,  {Cyr-Racine} F.-Y.,  {Pfrommer} C.,
  {Bringmann} T.,   {Sigurdson} K.,  2016, \mn@doi [\mnras]
  {10.1093/mnras/stw1076}, \href
  {https://ui.adsabs.harvard.edu/abs/2016MNRAS.460.1399V} {460, 1399}

\bibitem[\protect\citeauthoryear{{Voronoi}}{{Voronoi}}{1908}]{voronoi1908}
{Voronoi} G.~F.,  1908, J. Reine Angew. Math, 134, 198

\bibitem[\protect\citeauthoryear{{Walsh} \& {Tinker}}{{Walsh} \&
  {Tinker}}{2019}]{kt19}
{Walsh} K.,  {Tinker} J.,  2019, \mn@doi [\mnras] {10.1093/mnras/stz1351},
  \href {https://ui.adsabs.harvard.edu/abs/2019MNRAS.488..470W} {488, 470}

\bibitem[\protect\citeauthoryear{{Wang} \& {White}}{{Wang} \&
  {White}}{2007}]{Wang2007}
{Wang} J.,  {White} S.~D.~M.,  2007, \mn@doi [\mnras]
  {10.1111/j.1365-2966.2007.12053.x}, \href
  {http://adsabs.harvard.edu/abs/2007MNRAS.380...93W} {380, 93}

\bibitem[\protect\citeauthoryear{{Weaire}, {Kermode}  \& {Wejchert}}{{Weaire}
  et~al.}{1986}]{wkw86}
{Weaire} D.,  {Kermode} J.~P.,   {Wejchert} J.,  1986, \mn@doi [Philosophical
  Magazine, Part B] {10.1080/13642818608240647}, \href
  {https://ui.adsabs.harvard.edu/abs/1986PMagB..53L.101W} {53, L101}

\bibitem[\protect\citeauthoryear{{Wechsler}, {Zentner}, {Bullock}, {Kravtsov}
  \& {Allgood}}{{Wechsler} et~al.}{2006}]{wechsler+06}
{Wechsler} R.~H.,  {Zentner} A.~R.,  {Bullock} J.~S.,  {Kravtsov} A.~V.,
  {Allgood} B.,  2006, \mn@doi [\apj] {10.1086/507120}, \href
  {http://adsabs.harvard.edu/abs/2006ApJ...652...71W} {652, 71}

\bibitem[\protect\citeauthoryear{{White}}{{White}}{1979}]{white79a}
{White} S.~D.~M.,  1979, \mn@doi [\mnras] {10.1093/mnras/186.2.145}, \href
  {https://ui.adsabs.harvard.edu/abs/1979MNRAS.186..145W} {186, 145}

\bibitem[\protect\citeauthoryear{{Yang}, {Neyrinck}, {Arag{\'o}n-Calvo},
  {Falck}  \& {Silk}}{{Yang} et~al.}{2015}]{yang+15}
{Yang} L.~F.,  {Neyrinck} M.~C.,  {Arag{\'o}n-Calvo} M.~A.,  {Falck} B.,
  {Silk} J.,  2015, \mn@doi [\mnras] {10.1093/mnras/stv1087}, \href
  {https://ui.adsabs.harvard.edu/abs/2015MNRAS.451.3606Y} {451, 3606}

\bibitem[\protect\citeauthoryear{{Yoshioka} \& {Ikeuchi}}{{Yoshioka} \&
  {Ikeuchi}}{1989}]{yi89}
{Yoshioka} S.,  {Ikeuchi} S.,  1989, \mn@doi [\apj] {10.1086/167467}, \href
  {https://ui.adsabs.harvard.edu/abs/1989ApJ...341...16Y} {341, 16}

\bibitem[\protect\citeauthoryear{{Zehavi}, {Contreras}, {Padilla}, {Smith},
  {Baugh}  \& {Norberg}}{{Zehavi} et~al.}{2018}]{zehavi+18}
{Zehavi} I.,  {Contreras} S.,  {Padilla} N.,  {Smith} N.~J.,  {Baugh} C.~M.,
  {Norberg} P.,  2018, \mn@doi [\apj] {10.3847/1538-4357/aaa54a}, \href
  {http://adsabs.harvard.edu/abs/2018ApJ...853...84Z} {853, 84}

\bibitem[\protect\citeauthoryear{{Zentner}, {Hearin}  \& {van den
  Bosch}}{{Zentner} et~al.}{2014}]{zhv14}
{Zentner} A.~R.,  {Hearin} A.~P.,   {van den Bosch} F.~C.,  2014, \mn@doi
  [\mnras] {10.1093/mnras/stu1383}, \href
  {http://adsabs.harvard.edu/abs/2014MNRAS.443.3044Z} {443, 3044}

\bibitem[\protect\citeauthoryear{{Zentner}, {Hearin}, {van den Bosch}, {Lange}
  \& {Villarreal}}{{Zentner} et~al.}{2016}]{zentner+16}
{Zentner} A.~R.,  {Hearin} A.,  {van den Bosch} F.~C.,  {Lange} J.~U.,
  {Villarreal} A.,  2016, arXiv e-prints, \href
  {https://ui.adsabs.harvard.edu/\#abs/2016arXiv160607817Z} {p.
  arXiv:1606.07817}

\bibitem[\protect\citeauthoryear{{van de Weygaert}}{{van de
  Weygaert}}{1991}]{vdw91}
{van de Weygaert} R.,  1991, \mn@doi [\mnras] {10.1093/mnras/249.1.159}, \href
  {https://ui.adsabs.harvard.edu/abs/1991MNRAS.249..159V} {249, 159}

\bibitem[\protect\citeauthoryear{{van de Weygaert}}{{van de
  Weygaert}}{1994}]{vdw94}
{van de Weygaert} R.,  1994, \aap, \href
  {https://ui.adsabs.harvard.edu/abs/1994A&A...283..361V} {283, 361}

\bibitem[\protect\citeauthoryear{{van de Weygaert} \& {Icke}}{{van de Weygaert}
  \& {Icke}}{1989}]{vdwi89}
{van de Weygaert} R.,  {Icke} V.,  1989, \aap, \href
  {https://ui.adsabs.harvard.edu/abs/1989A&A...213....1V} {213, 1}

\bibitem[\protect\citeauthoryear{{van den Bosch} \& {Ogiya}}{{van den Bosch} \&
  {Ogiya}}{2018}]{vdbg18}
{van den Bosch} F.~C.,  {Ogiya} G.,  2018, \mn@doi [\mnras]
  {10.1093/mnras/sty084}, \href
  {https://ui.adsabs.harvard.edu/abs/2018MNRAS.475.4066V} {475, 4066}

\makeatother
\end{thebibliography}

\appendix
\section{Void probability function}
\label{app:negbin}
\noindent
The void probability function (VPF) of galaxies in large surveys and of dark matter haloes in $N$-body simulations has been very well studied in the literature. Such studies typically consider spherical regions whose volume $V$ is taken to be a control variable, as opposed to the integral over $V_U$ (equivalently, $\bar N$) which appears in \eqn{eq:<y^2>}. Empirically, one finds \citep[e.g.,][]{croton+04} that the galaxy VPF thus obtained from cosmological volumes is well-described by a universal form in terms of the scaling variable 
\be
\lambda \equiv n_{\rm trc}V\bar\xi_2(V)\,,
\label{eq:lambda-def}
\ee
with a shape given by the negative binomial form for the reduced VPF \citep{fry86,eg92}:
\be
\chi_{\textrm{NB}}(\lambda) = \frac1{\lambda}\,\ln\left(1+\lambda\right)\,.
\label{eq:negbin}
\ee
This universal form is well-motivated for the dark matter field whose correlation functions are reasonably well-described by the hierarchical clustering model \citep[see, e.g.,][]{ml-r87,vgph94}. The reason why the VPF is approximately universal for \emph{galaxies} (whose correlation hierarchy does not obey the hierarchical clustering model) was discussed by \citet{fc13} in the context of the halo model. Below, we use \eqn{eq:negbin} to interpret our numerical results for the VVF. 

\section{Sampling effects}
\label{app:downmaskassembly}

\begin{figure*}
\centering
\includegraphics[width=0.85\textwidth]{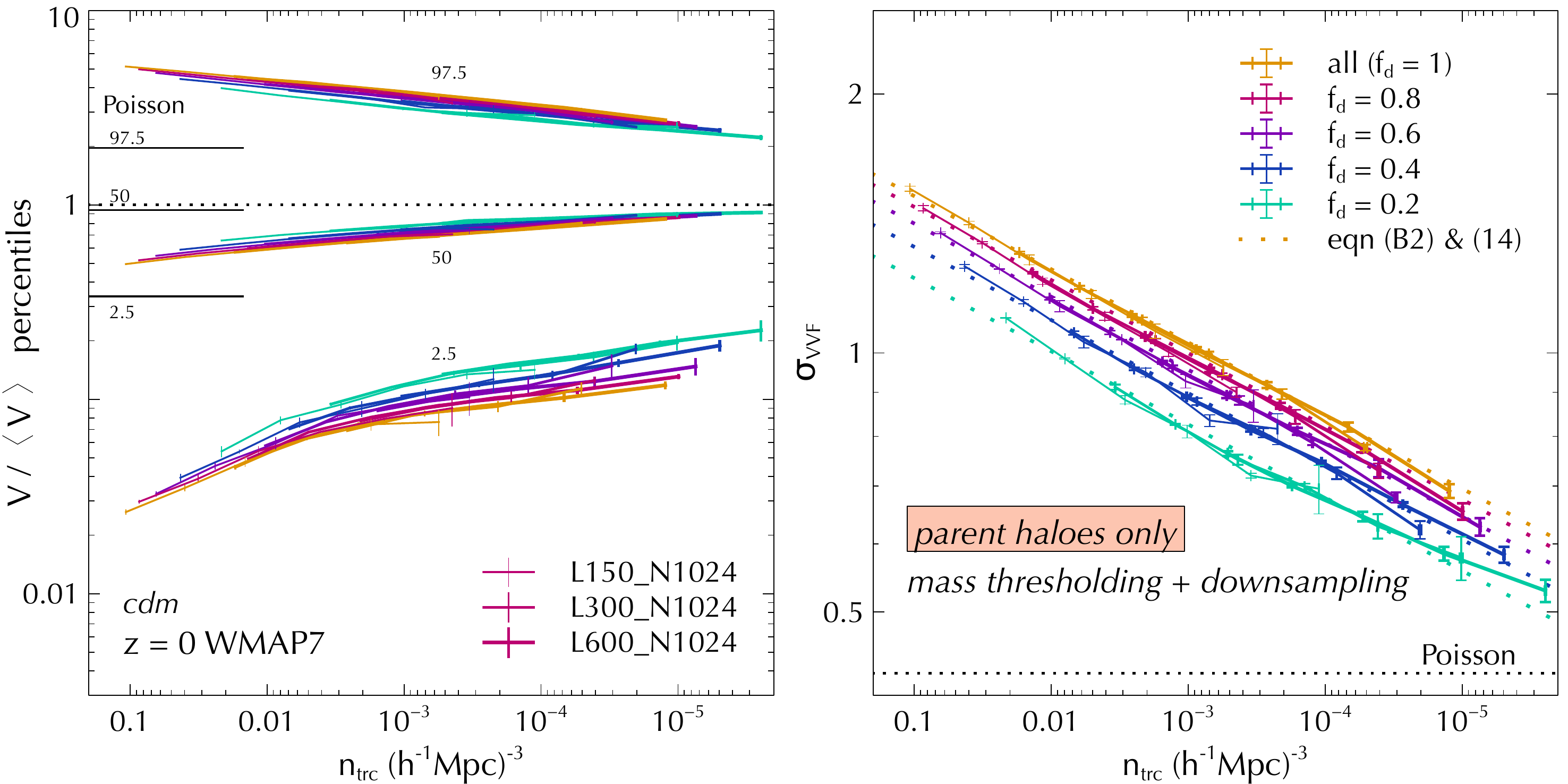}
\caption{{\bf Effects of downsampling:} 
VVF percentiles \emph{(left panel)} and standard deviation $\sigma_{\rm VVF}$ \emph{(right panel)} for real-space parent halo samples thresholded by mass $m_{\rm 200b}$, downsampled to $f_{\rm d}\times N_{\rm trc}$ where $N_{\rm trc}$ is the total number of tracers and $0<f_{\rm d}\leq1$ is a constant.
Values of $f_{\rm d}$ are colour-coded as indicated by the legend of the \emph{right panel}. 
Dotted lines in the \emph{right panel} show the fit from \eqn{eq:stdVVF-fit-ntrcfd}, with the full sample ($f_{\rm d}=1$) result given by \eqn{eq:stdVVF-fit-ntrc} and the dependence of the amplitude on $f_{\rm d}$ given by \eqn{eq:A(fd)}.
For comparison, percentiles and $\sigma_{\rm VVF}$ for Poisson distributed tracers are indicated by the horizontal line segments and the horizontal dotted line in the \emph{left} and \emph{right panels}, respectively.
Results are shown at $z=0$ in the WMAP7 CDM simulations for three configurations as indicated by the legend of the \emph{left panel} and averaged over all available realisations (see Table~\ref{tab:sims}), with error bars showing the standard deviation across the realisations.
As $f_{\rm d}$ decreases from unity to small values, $\sigma_{\rm VVF}$ and $y_{97.5}$ decrease while $y_{2.5}$ and the median $y_{50}$ increase, i.e., the VVF becomes narrower for smaller $f_{\rm d}$. See text for a discussion.
}
\label{fig:vvfpercstd-down}
\end{figure*}

\noindent
In this Appendix, we explore the dependence of the shape of the VVF on three aspects of sample selection: downsampling, masking and assembly bias. The first is physically motivated, from the point of view of scatter in the galaxy-dark matter connection: the selection of galaxies by observable properties such as luminosity or stellar mass is known to produce a downsampled version of the halo distribution, since not all haloes of a given mass contain galaxies of a given type. The second is observationally motivated, by the fact that realistic galaxy surveys typically do not cover contiguous patches of the sky due to the presence of, e.g., bright stars or other contaminants. The last can be a potential systematic for galaxy samples selected by imperfect mass proxies that correlate with other variables sensitive to the halo environment \citep[see, e.g.,][]{zhv14}. Since the VVF is intimately connected to spatial inter-connectivity of the sample under question, one must understand these aspects of sampling before drawing conclusions from any VVF analysis. 

We display results below for the WMAP7 CDM simulations at $z=0$, focusing on samples selected by a threshold on mass $m_{\rm 200b}$. We will compare results for parent haloes in real space with those of redshift-space samples containing substructure.

\begin{figure*}
\centering
\includegraphics[width=0.85\textwidth]{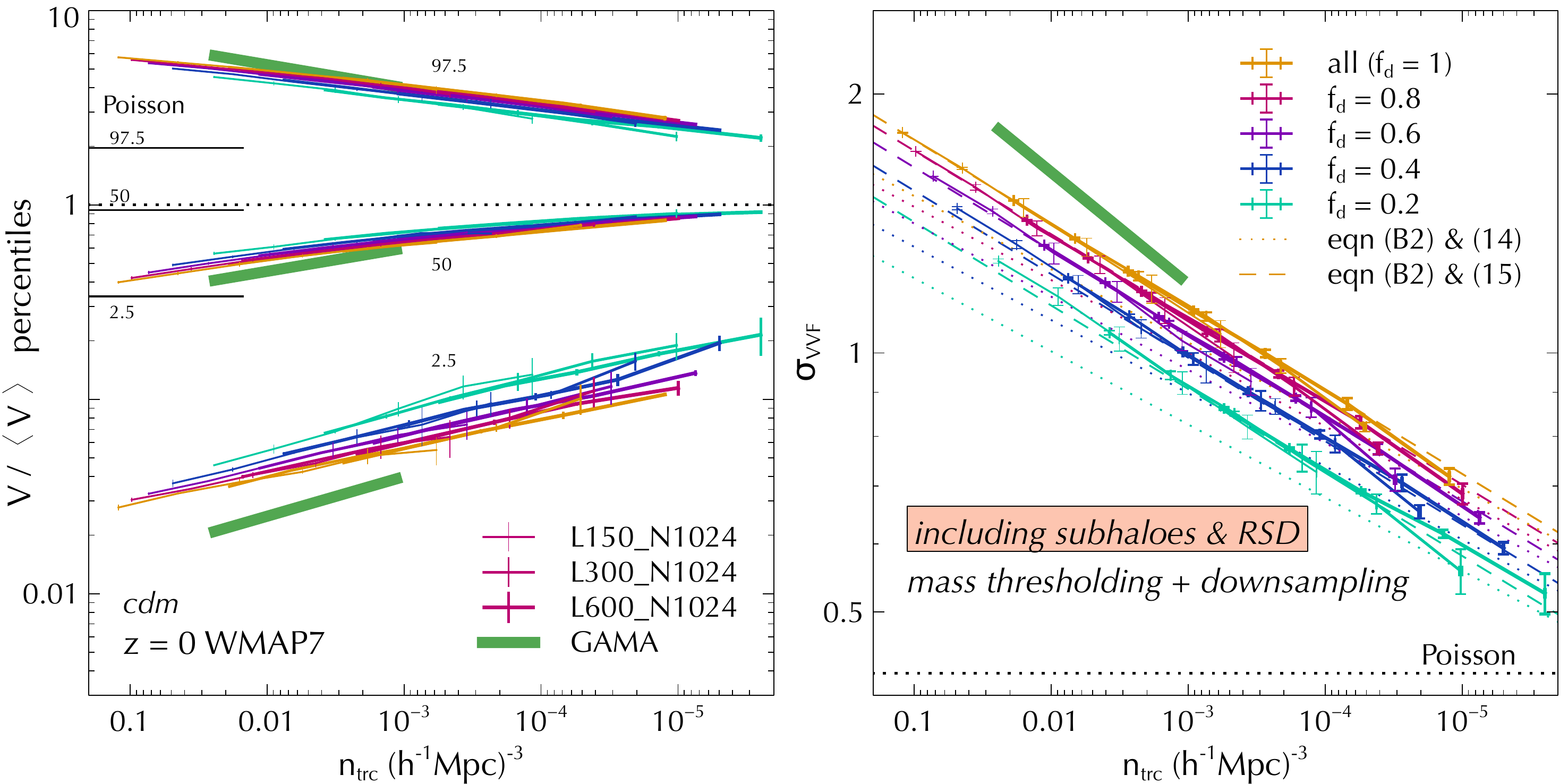}
\caption{{\bf Effects of downsampling:} Same as Figure~\ref{fig:vvfpercstd-down}, showing results including subhaloes with RSD. Thick solid green curves show the power-law fits to GAMA measurements from Table~\ref{tab:GAMA}. Dotted curves in the \emph{right panel} are repeated from Figure~\ref{fig:vvfpercstd-down}, while dashed curves show the fit in \eqn{eq:stdVVF-fit-ntrcfd} with the full sample result now given by \eqn{eq:stdVVF-fit-ntrc-RSDsub}. See text for more details.
}
\label{fig:vvfpercstd-down-RSDsub}
\end{figure*}

\subsection{Downsampling}
\label{app:down}
\noindent
To see why downsampling must affect the shape of the VVF, consider a sample of $N_{\rm trc}$ tracers that is uniformly downsampled by a factor $f_{\rm d}$, where $0<f_{\rm d}<1$. The number density of the sample decreases by this factor while keeping the total volume occupied by the sample unchanged at $V_{\rm tot}$. Moreover, such a random downsampling also leaves the hierarchy of correlation functions of the sample unchanged \citep{sheth96}. 
Using the negative binomial model for the VPF and assuming that the integral over $\bar N$ in \eqn{eq:<y^2>} has support over a range $\left(N_\ast-\Delta N/2,N_\ast+\Delta N/2\right)$ with $N_\ast\gg1$, it is possible to argue that, in the strong clustering regime, the difference between the variances of the VVF of the full sample and the downsampled tracers (denoted with a subscript `d') is approximately given by
\be
\avg{y^2}-\avg{y_{\rm d}^2} \simeq 1.179\left(1-f_{\rm d}\right)\,\avg{\Delta N}_{(z,\mu)} > 0\,,
\label{app:down-approxvardiff}
\ee
so that $\sigma_{\rm VVF}$ decreases as $f_{\rm d}$ is made smaller.

\begin{figure*}
\centering
\includegraphics[width=0.85\textwidth]{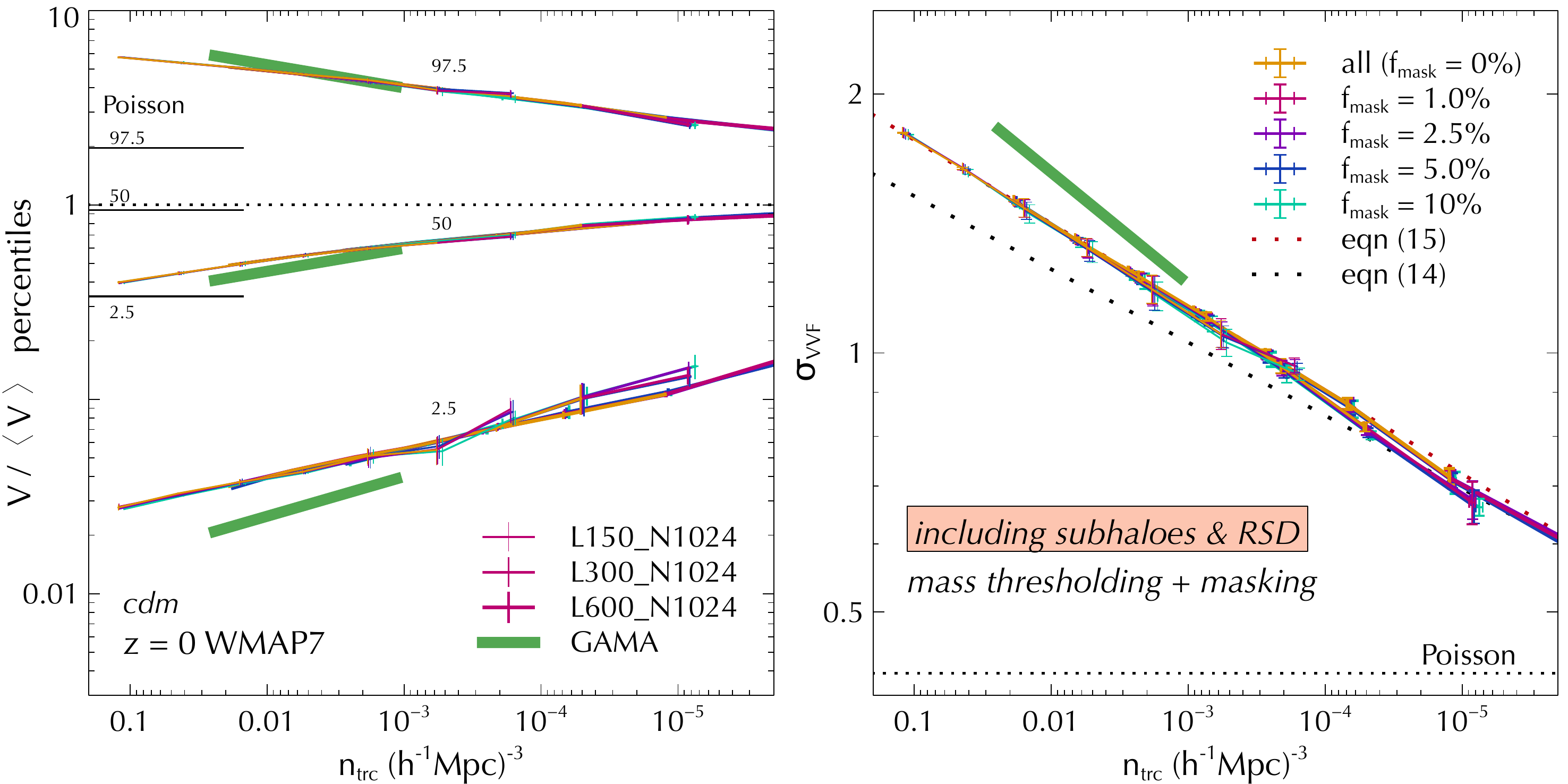}
\caption{{\bf Effects of masking:} 
Same as Figure~\ref{fig:vvfpercstd-down-RSDsub}, showing results for redshift-space samples containing subhaloes after masking a fraction $f_{\rm mask}$ of area perpendicular to the observer line-of-sight as described in the text. Evidently, masking has no significant effect on the VVF, even for extreme cases such as $f_{\rm mask}=10\%$. See text for a discussion.}
\label{fig:vvfpercstd-mask-RSDsub}
\end{figure*}

More intuitively, we can understand the decrease of $\sigma_{\rm VVF}$ of clustered tracers as follows. Upon downsampling, each tracer that `disappears' gives its volume to its neighbours. In a clustered sample, tracers with small volumes preferentially give their volume to other small-volume tracers. Also, conservation of total volume means there are many more small-volume tracers than large-volume tracers -- e.g., this is evident from the asymmetry of the VVF percentiles in, say, Figure~\ref{fig:vvfperc-ntrc}. Downsampling on the other hand does not discriminate between tracer type, so that most of the redistribution of volumes occurs at the small volume end, leaving the large volume end mostly unchanged. The VVF is therefore preferentially adjusted at small $y$ (being pushed towards larger $y$), thus decreasing the width of $p(y)$.

Figure~\ref{fig:vvfpercstd-down} shows the effects of uniformly downsampling a mass thresholded sample of parent haloes in real space by a factor $f_{\rm d}$. We see a clear trend in the \emph{left panel} for the entire distribution to become narrower as $f_{\rm d}$ decreases from unity to small values. Correspondingly, the standard deviation $\sigma_{\rm VVF}$ in the \emph{right panel} decreases with $f_{\rm d}$ as expected from the previous arguments. We have found that the results $\sigma_{\rm VVF}(n_{\rm trc},f_{\rm d})$ are nicely described (dotted lines) by the separable form
\be
\sigma_{\rm VVF}(n_{\rm trc},f_{\rm d}) = A(f_{\rm d}) \times \sigma_{\rm VVF}(n_{\rm trc})\,,
\label{eq:stdVVF-fit-ntrcfd}
\ee
where $\sigma_{\rm VVF}(n_{\rm trc})$ is given by \eqn{eq:stdVVF-fit-ntrc} and the amplitude $A(f_{\rm d})$ is fit by
\be
A(f_{\rm d}) = 1 - 0.1076\left(1- f_{\rm d}\right) - 0.1735 \left(1-f_{\rm d}\right)^{2}\,.
\label{eq:A(fd)}
\ee
Figure~\ref{fig:vvfpercstd-down-RSDsub} is formatted identically to Figure~\ref{fig:vvfpercstd-down} but shows results for redshift-space samples that included subhaloes. As in the case of the all-tracer sample, the downsampled VVFs for each value of $f_{\rm d}$ are broader than their real-space, parent-only counterparts. These redshift-space results including substructure are also reasonably well-described by the separable form \eqref{eq:stdVVF-fit-ntrcfd} with $\sigma_{\rm VVF}(n_{\rm trc})$ now given by \eqn{eq:stdVVF-fit-ntrc-RSDsub} and $A(f_{\rm d})$ again given by \eqn{eq:A(fd)}.
Such fits can be useful for halo occupation distribution (HOD) analyses involving mass-dependent downsampling of the halo population to describe galaxy populations thresholded by, say, luminosity or stellar mass.

\subsection{Masking}
\label{app:mask}
\noindent
To test for the effects of survey masks that include holes due to excluding bright stars, etc., we did the following. We pick the $z$-direction as the observer line of sight (same as that chosen for RSD) and randomly place a number of cubic masks on the $x$-$y$ plane, each of size approximately $2$ arcmin at $z\simeq0.5$. The number of such cubes is chosen such that the fraction of area being masked is fixed at some value $f_{\rm mask}$. We then exclude all tracers ``behind'' the masked region and proceed with the Voronoi tessellation. We generate randoms uniformly in the full survey area and then exclude those randoms ``behind'' the mask, as done for the tracers. The Voronoi volume of each unmasked tracer is calculated as discussed before, by counting the number of randoms assigned to it. Thus the masked volume is not counted in any Voronoi cell.

The resulting VVFs for various values of $f_{\rm mask}$ are shown in Figure~\ref{fig:vvfpercstd-mask-RSDsub}. We see that masking has \emph{no noticeable effect} on the VVF shape, even for extreme cases like $f_{\rm mask}=0.1$.

\subsection{Assembly bias}
\label{app:assembly}
\noindent
The fact that the VVF is intimately connected to halo clustering (c.f. section~\ref{subsubsec:b1}) means that samples selected by \emph{any} halo property that correlates with large-scale clustering would show non-trivial effects in their VVF. In fact, many such variables exist, since halo assembly is known to correlate tightly with the large-scale cosmic web environment \citep[the so-called halo assembly bias,][]{st04,gsw05,wechsler+06}. We consider these effects in this Appendix.

\begin{figure}
\centering
\includegraphics[width=0.45\textwidth]{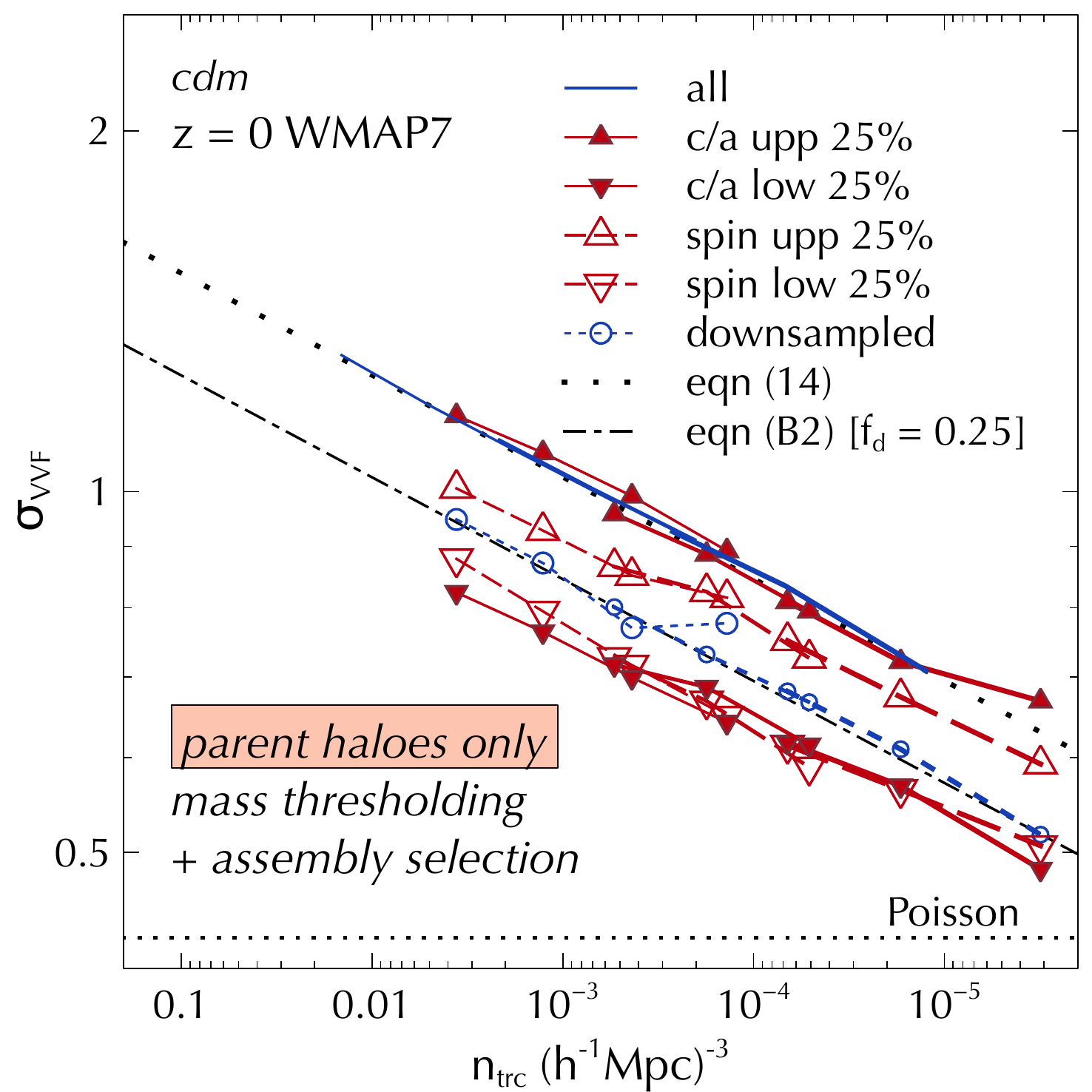}
\caption{{\bf Effect of halo assembly bias:} 
Standard deviation $\sigma_{\rm VVF}$ of the VVF of real-space parent halo samples selected by a mass threshold and further split into upper and lower quartiles (respectively, upward and downward pointing triangles) of the assembly variables $c/a$ (filled symbols) and spin (open symbols), as described in the text. Solid blue lines show the full sample result while open blue circles joined by a dotted line show $\sigma_{\rm VVF}$ for tracers randomly downsampled to one quarter of the full sample.
Dotted black curve shows the fit from \eqn{eq:stdVVF-fit-ntrc} while the dash-dotted black curve shows the fit from \eqn{eq:stdVVF-fit-ntrcfd} using \eqn{eq:stdVVF-fit-ntrc} and $f_{\rm d}=0.25$.
Results are shown at $z=0$ in the WMAP7 CDM simulations for one realisation each of the same three configurations as in Figure~\ref{fig:vvfpercstd-down}. See text for a discussion.
}
\label{fig:vvfstd-assemblybias}
\end{figure}

Figure~\ref{fig:vvfstd-assemblybias} shows $\sigma_{\rm VVF}$ for real-space mass-thresholded samples of parent haloes that are further split according to the values of the halo asphericity $c/a$ and dimensionless spin $\lambda$. 
The halo asphericity $c/a$ is the ratio of the smallest and largest eigenvalues of the weighted mass ellipsoid tensor of each halo, which is calculated by \textsc{rockstar} using the iterative algorithm prescribed by \citet{allgood+06}.
The dimensionless spin is calculated by \textsc{rockstar} using the bound particles of any halo as $\lambda \equiv J |E|^{1/2} / (G m^{5/2})$, where $J$, $E$ and $m$ are, respectively, the magnitude of the angular momentum, the total energy and mass of the halo \citep{peebles69}. 
For each of the variables $c/a$ and $\lambda$, we select haloes that lie in the upper or lower quartile of that variable in some narrow mass range. The Voronoi tessellations are then separately computed for the respective samples. We then repeat the exercise as a function of threshold mass $m_{\rm lim}$ and quote results as a function of the number density $n_{\rm trc}$ of each individual (sub)sample. For comparison, the open circles show the result of randomly downsampling the halo catalog for each $m_{\rm lim}$ by a factor $f_{\rm d}=0.25$; these are well-described by the fitting function in \eqn{eq:stdVVF-fit-ntrcfd}.

\emph{Relative to the downsampled curve}, the results for the upper and lower quartiles of the assembly variables show trends consistent with known results on assembly bias. In particular, halo populations with higher (lower) asphericity or spin have larger (smaller) values of $\sigma_{\rm VVF}$ than randomly sampled haloes with the same number density, consistent with the fact that these populations also have higher (lower) values of large-scale bias \citep[e.g.,][]{fw10,rphs19}. Figures~\ref{fig:vvfpercstd-down} and~\ref{fig:vvfstd-assemblybias} show, however, that the effects of downsampling and splitting by an assembly variable can be very degenerate (a case in point being the $\sigma_{\rm VVF}$ curve for the upper $c/a$ quartile which is, coincidentally, nearly identical to the all-halo result). This highlights the need for caution when interpreting the results of, both, analyses aimed at understanding halo assembly effects in galaxy evolution as well as cosmological analyses that might be contaminated by assembly effects.

\section{Linear bias estimator}
\label{app:hbyhbias}
\noindent
In this Appendix, we construct a least squares estimator suitable for the estimation of linear halo-by-halo bias which improves upon the estimator presented by \citet{phs18a}.

Consider a general situation in which a sample of $N_{\rm h}$ haloes in a volume $V$ is being used to estimate their mean linear bias $b_1$. We are eventually interested in the limit in which $N_{\rm h}=1$. Denote the measured halo-matter cross-power spectrum in bins of wavenumber $k$ as $P_\times(k)$, the corresponding matter auto-power spectrum as $P(k)$, and their respective variances as $\sigma_\times^2(k)$ and $\sigma^2(k)$. For simplicity, in the following we will assume that these measurements are uncorrelated, although this is not exactly true. We also suppress the $k$-dependence of individual terms below. It is then easy to show that the estimator
\be
\hat b_1 = \frac{\sum_kw_kP_\times}{\sum_kw_kP}
\label{app:b1-minvar}
\ee
minimises the statistic $\chi^2 = \sum_k\left(P_\times-b_1P\right)^2/\left(\sigma_\times^2+\sigma^2\right)$, provided the weights $w_k$ are given by
\be
w_k = P / \left(\sigma_\times^2+\sigma^2\right)\,.
\label{app:wts-minvar}
\ee
In the limit of Gaussian errors, \citet{smith09} shows that the variances $\sigma^2$ and $\sigma_\times^2$ can be approximated by
\begin{align}
\sigma^2 &\simeq \frac{2P^2}{N_k} + \Cal{O}\left(\frac1{N_{\rm dm}}\right) \notag\\
\sigma_\times^2 &\simeq \frac1{N_k}\left(P_{\rm hh}P+P_\times^2+\frac{PV}{N_{\rm h}}\right) + \Cal{O}\left(\frac1{N_{\rm dm}}\right)\,,
\label{app:Gaussvar}
\end{align}
where $N_k$ is the number of $k$ modes in the bin, $P_{\rm hh}$ is the halo auto-power spectrum and $N_{\rm dm}$ is the dark matter particle number.

In the limit where the halo sample is large, i.e. $N_{\rm h}\gg1$, we have 
\be
\sigma^2+\sigma_\times^2 \simeq \frac1{N_k}\left(P_\times^2+2P^2+P_{\rm hh}P\right) \simeq \frac{\#\,P^2}{N_k}\,,
\label{app:totvar-largeNh}
\ee
where $\#$ is a number of order unity related to the halo bias of the sample. This leads to $w_k = N_k/(\# P)$ so that 
\be
\hat b_1 = \frac{\sum_kN_kP_\times/P}{\sum_kN_k}\,,
\label{app:b1-phs18}
\ee
which is the same as used by \citet{phs18a}.

The situation at hand, however, has the opposite extreme of $N_{\rm h}=1$. In this case, we get
\be
\sigma^2+\sigma_\times^2 \simeq \frac1{N_k}\left(P_\times^2+2P^2+P_{\rm hh}P + PV\right) \simeq \frac{PV}{N_k}\,,
\label{app:totvar-hbyh}
\ee
where, in the last approximation, we used the fact that $P_{\rm hh}P,P_\times^2,P^2\ll PV$ for nearly all $k$ in typical cosmologies when $V\gtrsim(100\Mpch)^3$. This leads to $w_k = N_k/V$, so that
\be
\hat b_1 = \frac{\sum_kN_kP_\times}{\sum_kN_kP}\,,
\label{app:b1-final}
\ee
which is the estimator used in this work and which we advocate for all future analyses involving halo-by-halo bias. We have verified that using this estimator does not change any of the conclusions of \citet{phs18a} or \citet{rphs19}.

\label{lastpage}

\end{document}